\documentclass[11pt]{article}
\usepackage{amssymb, amsfonts, amsgen, amsbsy, mathrsfs}
\usepackage{graphicx}
\newcommand{\be}{\begin{equation}}
\newcommand{\ee}{\end{equation}}
\begin{document}
\def\theequation{\arabic{section}.\arabic{equation}}
\begin{titlepage}
\title{Evolving black hole horizons in General Relativity and 
alternative gravity}
\author{Valerio Faraoni \\ \\
{\small \it Physics Department and STAR Research Cluster} \\
{\small \it  Bishop's University, 
2600 College St., Sherbrooke, Qu\'{e}bec, Canada J1M~1Z7}}
\date{}
\maketitle   \thispagestyle{empty}  \vspace*{1truecm}

\begin{abstract} 
From the microscopic point of view, realistic black holes are 
time-dependent and the teleological concept of event horizon fails. At present, 
the apparent or the trapping horizon seem its best replacements in various 
areas of black hole physics. We discuss the known phenomenology of 
apparent and trapping horizons for analytical solutions of General Relativity 
and alternative theories of gravity. These specific examples (we focus on 
spherically symmetric inhomogeneities in a background cosmological 
spacetime) are useful as toy models for research on various aspects of  
black hole physics.
\end{abstract}
\vspace*{1truecm} 
\begin{center} Keywords: black holes, evolving horizons, apparent horizons.
\end{center}     
\end{titlepage}   \clearpage

\section{Introduction}

In the literature on gravitation and quantum field theory in curved space, 
several kinds of horizon are studied, including Rindler horizons for 
accelerated observers in Minkowski space, black hole horizons, and 
cosmological horizons. Research on  classical and semiclassical black hole 
physics has unveiled inner, outer, Cauchy, and extremal 
horizons. The early literature on black holes 
and the 1970s development of black hole thermodynamics  
focused on  stationary (or even static) black holes and on event 
horizons (see, {\em e.g.}, Refs.~\cite{Poissonbook, FrolovNovikov, Waldbook}). 
However, highly dynamical situations are of great interest for theorists, 
including gravitational collapse, the merger of a black hole with 
a compact object, evaporation of a small black hole 
due to Hawking radiation, and black 
holes  interacting with  non-trivial environments. Examples of nontrivial
environments occur in the case of black holes 
accreting or emitting  gravitating matter such as 
Vaidya spacetimes (which recur in mathematical studies of horizon dynamics); 
black holes immersed  in a cosmological background other than de Sitter 
space; black holes emitting (and possibly accreting) Hawking radiation in the 
final stages of their evolution when backreaction is significant; or black 
holes with variable mass due to other conceivable processes. 
In all these situations the concept of event horizon must be generalized. 
Moreover, if the black hole is located in a non-Minkowskian background, its 
mass-energy (which is usually the internal energy appearing
in the first law of black hole thermodynamics) needs to be defined 
carefully through some notion of quasi-local energy, and is 
related to the notion of horizon.

In Rindler's words, an horizon is ``a frontier between things 
observable and things unobservable'' \cite{Rindler56}. The horizon concept,
which is the product of strong gravity,  
is perhaps the most impressive feature of a black hole spacetime and, 
traditionally, the one that best characterizes the black hole concept itself. 
Various notions of black hole horizon studied in the technical literature 
include event,  Killing, inner, outer, Cauchy, apparent, trapping, 
quasi-local, isolated, dynamical, and slowly evolving 
horizons (for reviews see Refs.  
\cite{Poissonbook, Waldbook, Boothreview, Nielsenreview, AshtekarKrishnan,
GourghoulhonJaramillo08}).  
The notions of event, apparent, trapping, and dynamical horizon usually  coincide for 
stationary black holes but they are quite different from each other in the 
case of dynamical black holes with masses evolving in time. 

The usual definition of black hole event horizon turns out to be 
pretty much useless for practical purposes in highly dynamical situations 
because it requires the knowledge of the entire causal structure of spacetime 
(including future null infinity $\mathscr{I}^{+}$), which is physically  
impossible \cite{BenDov05, AshtekarGalloway05, Nielsenreview}. 
Time-evolving black holes are not just a theoretician's playground but 
they are also important for astrophysics. The remarkable improvements in 
astronomical techniques in recent years and their projected developments in 
the near future have stressed the important roles that stellar 
mass and supermassive black holes play in the modelling of astrophysical 
systems. The improvement of ground-based gravitational wave detectors 
(presumably nearing their first detection in the next few years) and the 
development of space-based detectors prompt enormous theoretical efforts to 
predict in detail the gravitational waves emitted by astrophysical black 
holes and build template banks for interferometric detectors. 
Progress in this theoretical programme went hand-in-hand with the 
improvement in  computing power; however, for numerical calculations on 
black hole systems, the notion of event horizon is  again of little use 
in the highly dynamical situations involving gravitational collapse or the 
evolution or merger of a close binary system with a black hole component. In 
practice, ``black holes'' are  identified with outermost marginally trapped 
surfaces and apparent horizons in numerical studies 
\cite{Thornburg07, BaumgarteShapiro03, ChuPfeifferCohen11, Booth08}.

The concept of horizon is not limited to black holes: there are also Rindler 
horizons for uniformly accelerated observers in 
Minkowski space and cosmological horizons. In cosmology, in addition to the  
particle and event horizons familiar from the standard literature on inflation 
\cite{KolbTurner90, Mukhanovbook, LiddleLythbook}, 
cosmological apparent and trapping horizons 
have also been studied more and more intensely in the recent 
past. Shortly after the discovery of Hawking radiation from stationary 
black holes \cite{Hawkingradiation1, Hawkingradiation2} and the 
completion of black hole thermodynamics, it was pointed out  
that the event horizon of de  Sitter space should also be 
attributed a temperature and an entropy \cite{GibbonsHawking77}. The region 
of de Sitter space below this horizon is static and the de Sitter  
horizon does not evolve, therefore the latter can be considered, to a certain 
extent, as a cosmological analogue of the Schwarzschild event and Killing 
horizon. In this logic, the analogue of time-dependent black hole 
horizons would be the time-dependent apparent and trapping horizons of 
Friedmann-Lema\^itre-Robertson-Walker (FLRW) spacetimes.

Many theoretical efforts went into generalizing the ``standard'' black 
hole thermodynamics for event horizons to moving horizons or horizon 
constructs different from event horizons \cite{Collins92, 
Hayward:1993wb, AshtekarKrishnan}. Thermodynamical studies of FLRW 
apparent horizons have also appeared. In principle, while it is 
reasonable that ``slowly moving'' horizons are meaningful from the 
thermodynamical point of view in some adiabatic approximation, it is not 
at all clear that fast-evolving horizons constitute thermodynamical 
systems and, if they do, they would most likely require non-equilibrium 
thermodynamics, as opposed to equilibrium thermodynamics, for their 
description. This feature would clearly complicate the study of highly 
time-dependent horizons.

The (now classical) thermodynamics of stationary black hole horizons does not 
make reference to the field equations of the gravitational theory and, 
therefore, black holes in theories of gravity alternative to General 
Relativity (GR) can be usefully studied as well (it is true, however, that 
some dependence on the action remains, for example the  entropy of  a stationary 
black hole horizon of area $A$ 
in scalar-tensor gravity with a Brans-Dicke-like scalar field $\phi$ 
is not $S=A/4G$ in units $c=\hbar=1$, but rather $S=\phi A/4$, which 
can be understood naively by noting 
that $\phi \sim 1/G$ plays the role of the inverse of the effective gravitational 
coupling in these theories --- see Ref.~\cite{myentropy} for details).  
In recent years there 
has been a  renewed interest in alternative theories of gravity with various 
motivations. First of all, the search for a quantum theory of gravity 
has generated considerable interest in low-energy effective 
actions, which  invariably contain ingredients foreign to Einstein's theory 
such as scalar fields coupled non-minimally to the curvature (which give 
a scalar-tensor nature to the gravitational theory), or higher derivative terms, or 
non-local terms. From this point of view, the question is not {\em if} but 
{\em when} gravity will deviate from GR \cite{Niayesh}. Further motivation 
for extending GR comes from  attempts (see \cite{SotiriouFaraoni10review, 
DeFeliceTsujikawa10, bookSalv} for reviews) to explain the current acceleration 
of the universe discovered with type Ia supernovae without invoking an {\em ad 
hoc} dark energy \cite{AmendolaTsujikawabook}. One should also mention 
attempts to replace the concept of dark matter at galactic and cluster scales by
modifying not just relativistic, but even Newtonian gravity, given that dark 
matter particles seem to elude direct detection (there is yet no agreement from 
the various groups on reports of possible signals). 

From the astrophysical point of view, if primordial black holes formed in the 
early universe, they would have had a scale comparable to the Hubble scale and
very dynamical horizons evolving on the Hubble time scale, and an important 
 question is how fast these black holes could accrete and grow.

\begin{figure}
\centering
\includegraphics[width=8.5cm]{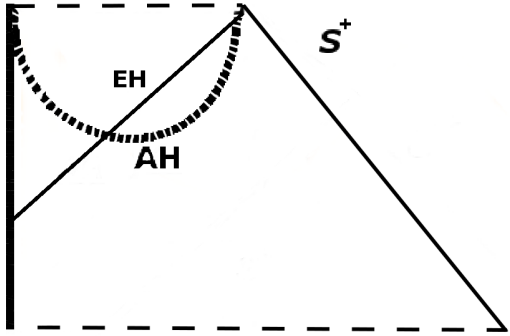}
\caption{The conformal diagram of an hypothetical cosmological black hole. 
The bottom horizontal (dashed) line represents a Big Bang singularity, the 
top horizontal line (dashed) is a spacelike black hole singularity, and an 
apparent horizon (marked AH) can change from timelike, to null, to 
spacelike, and it can be located inside or outside the event horizon 
(forty-five degrees line marked EH) according to whether the energy 
conditions are satisfied or not. If $R_{ab}l^al^b \geq 0$ for all null 
vectors $l^a$, then the apparent horizon lies inside the event horizon 
(\cite{Waldbook}, p.~311).
\label{Generic}} 
\end{figure}

To summarize, there are currently many avenues of research in theoretical  
gravitational physics in which evolving horizons play some 
role. Here we review the main definitions and properties of horizons and 
we focus on apparent and trapping  black hole  horizons. While it is 
questionable that apparent horizons are the ``correct'' notion of horizon 
in the dynamical case (and there are indications that they may not satisfy a 
quantum generalized second law \cite{Wall12}), they are 
the ones that are used in practice in numerical 
relativity and there seems to be no better candidate at the moment for the 
concept of ``horizon'' when time-dependence and interaction with the environment 
are allowed.  
Since only a few exact solutions of GR (and even 
less of other theories of gravity) are known for which the horizons are 
explicitly time-dependent, here we focus on solutions describing  black holes 
embedded in cosmological backgrounds, which have been studied in some detail. 
Fig.~\ref{Generic} shows how a conformal diagram of an hypothetical cosmological 
black hole may look like.

There are many additional reasons  for pursuing the study of analytical 
solutions of GR and of gravitational theories representing a central 
inhomogeneity in a cosmological background. The non-linearity of the field equations 
prevents the splitting of solutions into a ``background'' and a ``deviation'' from it in 
general, but we drop the quotation marks and from now on the term ``background'' refers 
to the asymptotic structure of spacetime. We have already mentioned the use of alternative theories of 
gravity, and of $f\left({R^c}_c\right)$ gravity in particular 
(where ${R^c}_c$ is the Ricci curvature of spacetime and $f$ is the Lagrangian 
density of the gravitational field), to explain the present 
acceleration of the universe without dark energy: since these theories 
are designed to produce a time-varying effective cosmological 
``constant'', black hole spacetimes in these theories are naturally 
asymptotically FLRW, not asymptotically flat, and are dynamical. It is also of
interest to study the spatial variation of fundamental constants throughout 
the universe, and scalar-tensor theories of gravity \cite{BransDicke61, Bergmann68, 
Wagoner70, Nordtvedt70} embody the
variation of the gravitational ``constant'', hence the search for analytical 
solutions describing condensations in cosmological spacetimes in these theories. 
Overall, very 
few such solutions are known in alternative gravity. But then one 
realizes that also analytical solutions of GR interpretable as central 
objects in cosmological backgrounds are quite interesting. The first 
solution of this kind is the McVittie spacetime \cite{McVittie}, which 
was invented to address the problem of whether, or how, the expansion of 
the universe affects local systems (see Ref. \cite{CarreraGiuliniRMD10} for a 
recent review on this subject). The McVittie solution of GR has a 
complex structure and is not yet completely understood 
\cite{Kleban, LakeAbdelqader11, Anderson11, Roshina1, Roshina2, 
AndresRoshina, SilvaFontaniniGuariento12}. 
Relatively few other solutions of GR with similar features 
have been discovered, including Swiss-cheese and other models 
\cite{Krasinskibook}.

Recent interest in cosmological condensations in GR arises also from a 
different attempt to explain the present cosmic acceleration without 
dark energy and without modifying gravity. The idea is that the 
backreaction of inhomogeneities on the expansion of the universe could be
 sufficient to produce the observed acceleration \cite{Buchert00, 
BuchertCarfora02, KolbMatarreseRiotto06, 
LarenaBuchertAlimi06, ParanjapeSingh07, LiSchwartz07, Wiltshire07, 
Wiltshire07PRL, Buchert08,LiSchwartz08, 
Larenaetal09}. However, the implementation of this idea has several 
formal problems and its proponents
 have not yet shown convincingly that this idea explains the magnitude 
or even the sign of the cosmic acceleration 
\cite{TsagasChallinorMaartens08, VitaglianoLiberatiFaraoni09} 
(more mathematically oriented work puts this proposed solution 
to the cosmic acceleration problem in jeopardy \cite{GreenWald11}). 
The study of exact inhomogeneous universes has also been pursued in 
yet another attempt to 
explain the current acceleration of the universe, the dominant idea 
being that we live inside a giant void which mimics an accelerated 
expansion; some of the analytical GR solutions considered are related to 
black holes in expanding universes (see \cite{BoleikoCelerier} for a 
review).

Independent motivation for the study of evolving horizons, and one not 
insignificant for astrophysics, comes from the renewed interest in exact 
models of spherical accretion by black holes, in particular the 
accretion of dark or phantom energy \cite{Babichevetal04, ChenJing05, 
IzquierdoPavon06, PachecoHorvath07, MaedaHaradaCarr08, 
GaoChenFaraoniShen08, Guarientoetal08, Guariento2, Guariento3, 
Guariento4, Guariento5, Guariento6, Sun08, Sun09, GonzalezGuzman09, 
HeWangWuLin09, Babichevetal11, Nouicer11, ChadburnGregory13}. 
This issue may again be 
relevant for primordial black holes which need to grow fast if they are 
to survive until the present era.

The plan of this review paper is the following: first, we review
 basic material in the next section. The following section discusses 
analytical solutions of GR, and is followed by a section on spacetimes with 
similar features in other theories of gravity. Due to space limitations it 
is not possible to discuss all the known solutions and their details, 
or to review all the works on cosmological black holes, but 
we do provide more detail for a few solutions to illustrate the techniques 
used, and the selection made is no doubt biased. Moreover, we do not 
discuss here the more mathematical approaches to time-evolving horizons and 
the various existence and uniqueness theorems
 for horizons.   
 The metric signature used is $-+++$ and we 
follow the conventions of Ref.~\cite{Waldbook} (the speed of light $c$ 
and Newton's constant $G$ are set to unity except where, occasionally, we restore 
them explicitly).

\section{Various notions of horizon}

Let us review briefly the geometry of the congruences of null geodesics 
crossing a horizon, which are used in the definition of non-stationary 
horizons.

\subsection{Null geodesic congruences and trapped surfaces}

Consider a congruence of null geodesics with tangent 
$l^a=dx^a/d\lambda$, where $\lambda$ is an affine parameter along each 
geodesic and $l_al^a=0$, $ l^c\nabla_c l^a=0$. The metric $ h_{ab}$ 
in the 2-space orthogonal to $l^a$ is determined by the following 
\cite{Poissonbook}: pick another null vector field $n^a$ such that 
$n_cn^c=0$ and $l^c n_c=-1$, then we have 
\be 
h_{ab} \equiv g_{ab}+l_a 
n_b +l_b n_a \,. \ee 
$ h_{ab}$ is purely spatial and ${h^a}_b$ is a 
projection operator on the 2-space orthogonal to $l^a$, {\em i.e.}, 
$
 h_{ab} \, l^a  =  h_{ab} \, l^b=0 \,, \;\;\;\;
{h^a}_a  =  2 \,,\;\;\;\;
{h^a}_c \, {h^c}_b =  {h^a}_b $. 
The choice of $n^a$ is not unique but the geometric quantities
of interest to us do not depend on it once $l^a$ is fixed. 
Let $\eta^a$ be the geodesic deviation vector 
(it corresponds to a one parameter subfamily of the congruence since its 
choice is not unique) and define the tensor field \cite{Waldbook, 
Poissonbook} 
\be 
B_{ab} \equiv \nabla_b \, l_a \,, 
\ee 
which 
satisfies $ l^b \nabla_b \, \eta^a= {B^a}_b \, \eta^b$ and is orthogonal 
to the null geodesics, $B_{ab}l^a=B_{ab}l^b=0$. The transverse part of 
the deviation vector is 
\be 
\tilde{\eta}^a \equiv {h^a}_b \, \eta^b=\eta^a+( 
n^c\eta_c) l^a 
\ee 
and the orthogonal component of $l^c\nabla_c \eta^a$, 
denoted by a tilde, is \cite{Poissonbook} 
\be 
\widetilde{ \left( l^c 
\nabla_c \eta^a \right)}= {h^a}_b {h^c}_d {B^b}_c \, \tilde{\eta}^d 
\equiv { \tilde{ B^a }}_d \, \tilde{\eta}^d \,. 
\ee 
The transverse 
tensor $ \widetilde{B}_{ab}$ is decomposed into its symmetric and 
antisymmetric parts, and the symmetric part is further decomposed into 
its trace and trace-free parts as \cite{Waldbook, Poissonbook} 
\be 
\widetilde{B}_{ab}=\widetilde{B}_{(ab)} + \widetilde{B}_{[ab]}= \left( 
\frac{\theta}{2} \, h_{ab}+\sigma_{ab} \right) + \omega_{ab} \,, 
\ee 
where the trace 
\be 
\theta= g^{ab} \widetilde{B}_{ab}= g^{ab} B_{ab} 
=\nabla_c \, l^c 
\ee 
is the expansion of the affinely parametrized congruence, 
\be 
\theta_{ab}= \frac{\theta}{2} \, h_{ab} 
\ee 
is the expansion tensor, 
\be 
\sigma_{ab}= \tilde{B}_{(ab)} -\frac{\theta}{2} \, h_{ab} 
\ee 
is the shear tensor, and 
\be 
\omega_{ab}= \tilde{B}_{[ab]} 
\ee 
is the vorticity tensor. The expansion, shear, and vorticity 
tensors are purely transversal ({\em i.e.}, orthogonal to $l^a$) and the 
shear and vorticity are trace-free. The shear scalar and 
vorticity scalar
\be 
\sigma^2 \equiv \sigma_{ab} \, 
\sigma^{ab} \,, \;\;\;\;\;\;\;\;\; \omega^2 \equiv \omega_{ab} \, 
\omega^{ab} 
\ee 
are non-negative. The expansion propagates along a null geodesic 
according to the celebrated Raychaudhuri equation 
\cite{Waldbook, Poissonbook}, which was the main tool in the proof 
of the singularity theorems of Hawking and Penrose 
\cite{HawkingEllis73, Waldbook},
\be 
\frac{d\theta}{d\lambda} = 
-\frac{\theta^2}{2} -\sigma^2+\omega^2 -R_{ab}l^a l^b \,; 
\ee 
similar propagation equations hold for $\sigma_{ab}$ and $\omega_{ab}$ 
\cite{Waldbook}. If the congruence of null geodesics with tangent $l^a$ 
is not affinely parametrized, the geodesic equation assumes 
the form
\be
l^c \nabla_c l^a = \kappa \, l^a \,,
\ee
where the 
quantity $\kappa$ which  measures the  failure of $l^{a}$ 
to be affinely parametrized is sometimes used,  on a horizon,  
as a possible  definition of surface gravity  
\cite{Nielsen:2007ac, PielahnKunstatterNielsen11}
(there are various inequivalent definitions of surface gravity 
in the literature). The 
expansion is now 
\be
\theta=\nabla_c \, l^c- \kappa 
\ee
or
\be \label{expansion} 
\theta_{l} = h^{ab}\nabla_{a}l_{b} = \left[ g^{ab} + 
\frac{l^ a n^b    +  n^a l^b}{\left( 
-n^{c}l^{d}g_{cd}\right)} \right] \nabla_{a}l_{b} \,.
\ee 
Eq. (\ref{expansion}) is independent  of the field equations of the theory 
 and can be applied when $l^c$ and $n^c$ are not 
normalized to satisfy $l^c n_c=-1$ as usual \cite{Nielsen:2007ac}.
With non-affine parametrization, the Raychaudhuri equation picks up an 
extra term \cite{Poissonbook},
\be
\frac{d\theta}{d\lambda} =\kappa \, 
\theta-\frac{\theta^2}{2} 
-\sigma^2+\omega^2 -R_{ab}l^a l^b \,.
\ee
A compact and orientable 2-surface embedded in 4-space has two independent  
directions orthogonal to it, corresponding to ingoing and 
outgoing  null rays.  One is naturally led to study congruences of 
ingoing and outgoing null geodesics with 
tangent fields  $l^a$ and $n^a$, respectively, and the way they propagate
in strong gravity.

Let us provide now some basic definitions for closed 2-surfaces 
\cite{AshtekarGalloway05, Boothreview, Nielsenreview, 
BoothBritsGonzalezVDB, GourghoulhonJaramillo08} 
(usually it is assumed that these 2-surfaces are spacelike 
\cite{AshtekarGalloway05, BenDov07, Boothreview} but we will not impose this 
requirement here):

\begin{itemize}
\item A {\em normal surface} corresponds to $\theta_{l} > 0$ 
and $\theta_{n} < 0$ (for example, a 2-sphere in Minkowski space 
satisfies this property).

\item A {\em trapped surface} \cite{Penrose65} corresponds to $\theta_{l}<0$ 
and $\theta_{n}<0$. The outgoing, in addition to the ingoing, 
future-directed null rays converge here instead of diverging and  
outward-propagating light is dragged back by strong gravity.

\item A {\em
 marginally outer trapped} (or {\em 
marginal}) {\em  surface (MOTS)} corresponds  to 
$\theta_{l} =0$ (where $l^a$ is the outgoing null normal 
to the surface) and $\theta_{n} < 0$.

\item An {\em untrapped  surface} is one with  
$\theta_{l} \theta_{n} < 0$.

\item An {\em antitrapped surface} corresponds to 
$\theta_{l} >0$ and $\theta_{n} > 0$ (both outgoing and 
ingoing future-directed null rays are diverging).

\item A {\em marginally outer trapped tube (MOTT)} is a 
3-dimensional surface which can be foliated entirely by 
marginally outer trapped (2-dimensional) surfaces. 

\end{itemize}

In GR Penrose has proved that, if a spacetime contains a trapped surface, 
the null energy condition holds, and there is a non-compact Cauchy surface 
for the spacetime, then this spacetime contains a singularity 
\cite{Penrose65}. Trapped surfaces are probably essential features in the 
concept of black hole and notions of ``horizon'' of practical utility will be 
identified with boundaries of spacetime regions containing trapped 
surfaces.  At present, the mathematical conditions for the existence and 
uniqueness of MOTSs are not completely clear. It is known that, 
in general, a MOTT can be distorted smoothly, hence 
MOTTs are non-unique \cite{Eardley98, AnderssonMarsSimon05, Boothreview}.

Let us review the various kinds of horizons appearing in 
the  literature on black holes, cosmology,  quantum field 
theory in curved spaces, and the corresponding 
thermodynamics.

\subsection{Event horizons}

The event horizon is the traditional notion of horizon for stationary black 
holes in GR.  An {\em event horizon} is { a connected component of the 
boundary $\partial \left( J^{-}( \mathscr{I}^+ )\right)$ of the causal past $ 
J^{-}( \mathscr{I}^+ )$ of future null infinity $\mathscr{I}^+ $ 
\cite{Hawking72, HawkingEllis73, Waldbook, Poissonbook}. This is the most 
peculiar feature of a black hole: the horizon is a causal boundary separating 
a region from which nothing can come out to reach a distant observer from a 
region in which signals can be sent out and eventually arrive to this 
observer. An event horizon is generated by the null geodesics which fail to 
reach future null infinity and, therefore (provided that it is smooth) is always a 
null hypersurface.

In astrophysics the concept of event horizon is implicitly taken as a 
synonym of black hole. However, since to define and locate an event 
horizon one must know all the future history of spacetime (one must know 
all the geodesics which do reach future null infinity and, tracing them back, 
the boundary of the region from which they originate), an event horizon 
is a globally defined concept. For an observer to state that a black 
hole event horizon has formed requires knowledge of the spacetime 
outside his or her future light cone, which is impossible to achieve 
unless the spacetime is stationary, the black hole has existed forever, 
and nothing changes (a common expression is that the event horizon has a 
teleological nature). It has been shown \cite{AshtekarKrishnan, 
BenDov07} that, in a collapsing Vaidya spacetime, an event horizon forms 
and grows starting from the centre and an observer can cross it and be 
unaware of it even though his or her causal past consists entirely of a 
portion of Minkowski space: the event horizon cannot be detected by this 
observer with a physical experiment. In other words, the event horizon 
``knows'' about events belonging to a spacetime region very far away and 
in its future but not causally connected to it 
 (``clarvoyance'' \cite{Boothreview, BengtssonSenovilla11, Bengtsson11}).

Due to its global nature, an event horizon is not a practical concept and it 
is nearly impossible to locate precisely an event horizon in a dynamical 
situation. Realistic astrophysical black holes have not existed forever but 
are formed by  gravitational collapse. In numerical relativity, codes 
designed to follow a collapse situation, a binary system merger, or other 
dynamical situations generating black holes, eventually crash and it is 
impossible to follow the evolution of a system to future null infinity. It is 
routine in numerical relativity to employ marginally trapped surfaces as 
proxies for event horizons ({\em e.g.}, \cite{BaumgarteShapiro03, 
ChuPfeifferCohen11, Booth08}).

The event horizon ${\cal H}$ is a tube in spacetime; a very common abuse 
of terminology consists of referring to the intersections of ${\cal H}$ 
with surfaces of constant time (which produce 2-surfaces) as ``event 
horizons'' (this abuse of terminology extends to all the notions of 
horizon that we define below).

\subsection{Killing  horizons}

When present, a Killing vector field $k^a$ satisfying the Killing equation $ 
\nabla_a k_b+\nabla_b k_a=0 $ defines a {\em Killing horizon} ${\cal H}$ of 
the spacetime $\left( M, g_{ab} \right)$, which is {\em a null hypersurface 
which is everywhere tangent to a Killing vector field $k^a$ which becomes 
null, $k^c k_c=0$, on ${\cal H}$}. This Killing vector field is timelike, $k^c 
k_c <0$, in a spacetime region which has ${\cal H}$ as boundary. Stationary 
event horizons in GR are Killing horizons \cite{Chrusciel96}, for 
example in the Schwarzschild geometry the event horizon $R=2M$ is also a 
Killing horizon and the timelike Killing vector $k^a=\left( \partial 
/\partial t \right)^a$ in the $R>2M$ region outside the event horizon becomes 
null at $ R=2M$ and spacelike for $R<2M$. An event 
horizon in a locally static spacetime is also a Killing horizon for the 
Killing vector $k^a=\left( \partial /\partial t \right)^a$ associated with 
the time symmetry \cite{Poissonbook}. If the spacetime is stationary and asymptotically flat 
(but not necessarily static), it must be axisymmetric and an event horizon 
is a Killing horizon for the Killing vector 
\be
 k^a =\left( \partial /\partial t \right)^a  +\Omega_H 
\left( \partial /\partial \varphi \right)^a \,,
\ee
which is a linear combination of the vectors associated with time 
and rotational symmetries, and where $\Omega_H$ is the angular 
velocity at the horizon (this statement requires the assumption 
that the Einstein-Maxwell equations hold and some assumption on 
the matter  stress-energy tensor \cite{HawkingEllis73, 
WaldLivRev}). When present, a Killing horizon defines a notion of 
surface gravity $\kappa_{{\small {\sf Killing}}}$, as we will see below.

Of course, the concept of Killing horizon is useless in spacetimes which do not 
admit timelike Killing vectors. Attempts to use conformal Killing horizons in 
spacetimes conformal to the Schwarzschild one (\cite{DyerHonig79, 
SultanaDyer04}, see also \cite{SultanaDyer05, McClureDyer06CQG, 
McClureAndersonBardahl07, McClureAndersonBardahl08, McClurethesis}) have not been fruitful. 
However, the introduction of the Kodama vector, which is defined in 
spacetimes without Killing vectors, in place of a Killing field is much 
more useful to introduce a  surface gravity.

\subsection{Apparent horizons}

A {\em future apparent horizon} is {\em the closure of a surface} (usually a 
3-surface) {\em which is foliated by marginal surfaces}; it is defined by the 
conditions on the time slicings \cite{Hayward:1993wb} 
\begin{eqnarray} 
& \theta_{l} = 0 \,,& \label{AHcondition1} \\ 
& \theta_{n} < 0 \,,& 
\label{AHcondition2} 
\end{eqnarray} 
where $\theta_l$ and $\theta_n$ are the 
expansions of the future-directed outgoing and ingoing null geodesic 
congruences, respectively (this more practical definition differs from that 
of Hawking and Ellis \cite{HawkingEllis73}, which is rather impractical 
\cite{Boothreview}).  Eq.~(\ref{AHcondition1}) expresses the fact that the 
congruence of future-pointing outgoing null rays momentarily stops expanding 
and, presumably, these rays turn around at the horizon, 
while the condition (\ref{AHcondition2}) originally served the purpose of 
distinguishing between black holes and white holes.

Apparent horizons are defined quasi-locally   
and are independent of the global causal structure of 
spacetime, contrary to 
event horizons. However, apparent horizons (and also trapping 
horizons, see below) depend on the choice of the foliation 
of the 3-surface with marginal surfaces \cite{WaldIyer91PRD, SchnetterKrishnan06} 
and, of  course, 
also the ingoing and outgoing null geodesics  orthogonal 
to these surfaces do, as well as their expansions  
$\theta_l$ and $\theta_n$  
\cite{FiguerasHubenyRangamaniRoss09}.  While the  
expansions are scalars, and are therefore independent  of 
the coordinate system chosen, sometimes a  coordinate choice 
 makes it easier to specify locally the 
foliation (for example by choosing spacelike surfaces  
of constant time coordinate --- different time coordinates 
identify different families of hypersurfaces of constant 
time), which is a geometric  object and is 
coordinate-independent. Congruences of outgoing 
and  ingoing null geodesics orthogonal to these surfaces 
will, of  course, change when changing the foliation. 
The dependence of apparent 
horizons on the spacetime slicing is  illustrated by the 
fact that non-symmetric slicings of the  Schwarzschild 
spacetime can be found for which  no apparent 
horizons exist \cite{WaldIyer91PRD, SchnetterKrishnan06}.

Apparent horizons are, in general, quite distinct 
from event horizons: for example, event and apparent horizons do not 
coincide in the Reissner-Nordstr\"om black hole (inner horizon) and in the  
Vaidya spacetime  
\cite{Poissonbook}. Also in static 
black holes  which are perturbed, the apparent and the event 
horizons do not coincide \cite{KavanaghBooth06}. During the spherical collapse of  
uncharged matter an event horizon forms before the 
apparent horizon does and the two come closer and closer 
until they eventually coincide asymptotically as the final 
static state is reached \cite{HawkingEllis73}. 

In GR, a black hole apparent horizon lies inside the event 
horizon provided that the null  curvature  
condition $R_{ab}\, l^al^b \geq 0 $~$\forall $~null 
vector~$l^a$   is satisfied \cite{HawkingEllis73}. This 
requirement coincides  with the null energy condition $ 
T_{ab}\l^al^b  \geq 0 $~$\forall $~null vector~$l^a$  if 
the Einstein equations are imposed, and in this case it 
is believed to be a reasonable condition on classical  
matter. However, Hawking radiation  itself violates the 
weak and the null energy conditions \cite{Visser96PRD}, as
does quantum matter. 
A simple scalar field non-minimally coupled to the curvature  
can also violate all of the energy conditions. The null 
curvature condition is easily violated also in alternative 
theories of gravity (for example, Brans-Dicke  
\cite{BransDicke61} and scalar-tensor \cite{Bergmann68, 
Wagoner70, Nordtvedt70} gravity) 
and the black hole apparent horizon  
has been observed to lie outside of the event  horizon during 
spherical collapse in Brans-Dicke theory,  although  it 
eventually settles inside of it when the static 
Schwarzschild state is achieved \cite{ScheelShapiroTeukolsky} (note that 
the GR black hole is the endpoint of collapse in general scalar-tensor gravity
for asymptotically flat black holes without matter other than the Brans-Dicke 
scalar field outside the horizon 
\cite{Hawking?, SotiriouFaraoniPRL}). 

To summarize, the cherished notion of event horizon seems rather useless 
in general dynamical situations and apparent horizons appear to be more 
practical in spite of their fundamental limitations of depending on the 
spacetime slicing and of possibly being timelike surfaces (this last 
drawback is probably the most puzzling one \cite{BoothBritsGonzalezVDB}).

\subsection{Trapping horizons}

A {\em future outer trapping horizon (FOTH)} is {\em the closure of a 
surface} (usually a 3-surface) {\em foliated by marginal surfaces such 
that on its 2-dimensional ``time slicings''} (\cite{Hayward:1993wb}, see 
also \cite{NielsenVisser06} and references therein) 
\begin{eqnarray} 
& \theta_{l} = 0 \,,& \label{THcondition1} \\ 
&&\nonumber\\ 
& \theta_{n} < 0 \,,& \label{THcondition2} \\ 
&&\nonumber\\ 
& {\cal L}_n \, \theta_{l} 
= n^{a}\nabla_{a} \, \theta_{l} < 0 \,, 
& \label{THcondition3} 
\end{eqnarray} 
where $\theta_l$ and $\theta_n$ are the expansions of the 
future-directed outgoing and ingoing null geodesic congruences, 
respectively.  The inequality (\ref{THcondition3}) serves the purpose of 
distinguishing between inner and outer horizons, {\em e.g.}, in the 
non-extremal Reissner-Nordstr\"om solution, and also distinguishes 
between apparent horizons and trapping horizons (it is not imposed for 
apparent horizons but it is required for trapping ones), and its sign 
distinguishes between future and past horizons.

The definition of a {\em past inner trapping horizon (PITH)} is obtained 
by exchanging $ l^a $ with $n^a $ and reversing the signs of the 
inequalities, \begin{eqnarray} & \theta_{n} = 0 \,,& 
\label{PITHcondition1} \\ &&\nonumber\\ & \theta_{l} > 0 \,,& 
\label{PITHcondition2} \\ &&\nonumber\\ & {\cal L}_l \theta_n = 
l^{a}\nabla_{a} \, \theta_{n} > 0 \,.& \label{PITHcondition3} 
\end{eqnarray} The past inner trapping horizon identifies a white hole 
or a cosmological horizon. As one moves just inside an outer trapping 
horizon, one encounters trapped surfaces, while trapped surfaces are 
encountered as as one moves just outside an inner trapping horizon.

As an  example, consider the static Reissner-Nordstr\"om 
black hole with the natural spherically symmetric foliation: the event 
horizon $r=r_{+}$ is a future outer trapping horizon (FOTH), the 
inner (Cauchy) horizon $r=r_{-}$ is a future inner trapping horizon (FITH), 
while the white hole horizons are past trapping horizons (PTHs).

Black hole trapping horizons have been associated 
with thermodynamics, and it is claimed that it is the 
trapping horizon area and not the area of the event 
horizon which should be associated with entropy in black 
hole thermodynamics \cite{Haijcek87, Hiscock89, 
Collins92, Nielsenreview}. This claim is controversial   
\cite{Sorkin97, CorichiSudarsky02, 
NielsenFirouzjaee12}.  The Parikh-Wilczek ``tunneling''  
approach \cite{ParikhWilczek00} is in principle applicable  
also to apparent and trapping horizons, not only to  
event horizons \cite{Visser03IJMPD, 
DiCriscienzoNadaliniVanzoZerbiniZoccatelli07, 
Clifton0804.2635, NielsenYeom09, JangFengPeng09, 
Nielsen0809.1711, AnghebenNadaliniVanzoZerbini05, 
HaywardDiCriscienzoNadaliniVanzoZerbini09} but also this 
aspect is not entirely free of controversy 
\cite{BarceloLiberatiSonegoVisser06}.

In general, trapping horizons do not coincide with event 
horizons. Dramatic examples are spacetimes which possess 
trapping 
horizons but not event horizons  \cite{RomanBergmann83, 
Hayward06PRL}. The difference between the areas of the trapping 
and the event horizon in particular spacetimes have been studied 
in Ref.~\cite{Nielsen10CQGareas}.

\subsection{Isolated, dynamical, and slowly evolving horizons}

Isolated horizons correspond   
to isolated systems in thermal equilibrium not interacting 
with their surroundings, which are described by a stress-energy 
tensor $T_{ab}$.   
The concept of isolated horizon has been introduced in 
relation with loop quantum gravity 
\cite{AshtekarBeetleFairhurst99, 
AshtekarBeetleDreyerFairhurstKrishnanLewandowskiWisnieski00, 
AshtekarCorichiKrasnov00,
AshtekarBeetleFairhurst00,
AshtekarCorichi00, 
FairhurstKrishnan01, 
AshtekarBeetleLewandowski02, 
AshtekarBeetleLewandowski02b}  and, in a general 
perspective, it is too restrictive when one  wants to 
allow mass-energy to cross the ``horizon'' (whichever way 
the latter is defined) in one direction or the other.

A {\em weakly isolated horizon} is {\em a null surface 
${\cal H} $ with null normal $l^a$ such that $\theta_l=0$, 
$-T_{ab}l^a$ is a future-oriented and causal vector, and ${\cal 
L}_l \left( n_b \nabla_a l^b \right)   =0$}. In this 
context $l^a$ is a Killing vector for the intrinsic 
geometry on  ${\cal H}$, without reference to the 
surroundings, and can therefore be used to define a 
``completely local Killing horizon'' when there are no 
energy flows across ${\cal H}$. The vector field $l^a$ 
generates a congruence of null geodesics on ${\cal H}$, 
which can be used to define a surface gravity $\kappa$ 
via the (non-affinely parametrized) geodesic equation
\be
l^a\nabla_a l^b=\kappa \, l^b \,,
\ee
which gives
\be
\kappa=-n_b l^a\nabla_a l^b 
\ee
using $n_b l^b=-1$. 
This surface gravity $\kappa$ is constant on the weakly 
isolated horizon ${\cal H}$, which corresponds to the 
zeroth law of thermodynamics. Since the vector field $n^a$ 
is not unique also this surface gravity is not unique. 

A Hamiltonian analysis of the phase space of isolated 
horizons, identifying boundary terms with the energies of 
these boundaries, leads to a first law of thermodynamics 
for isolated horizons with rotational symmetry 
\cite{AshtekarCorichiKrasnov00},
\be
\delta H_{ {\cal H}} = \frac{\kappa}{8\pi} \, \delta A 
+\Omega_{ {\cal H}} \delta J \,,
\ee
where $J$ is the angular momentum, $H_{{\cal H}}$ the 
Hamiltonian, $A$  the area of the 2-dimensional 
cross-sections of ${\cal H}$, and 
$\Omega_{ {\cal H}} $ 
the angular velocity of the horizon.

A {\em dynamical horizon} \cite{AshtekarKrishnan} is {\em a spacelike 
marginally trapped tube foliated by 
marginally trapped 2-surfaces} (MTT). This definition allows for energy 
fluxes across the dynamical horizon. A set of flux laws 
describing the related changes in the area of the dynamical 
horizons have been formulated \cite{AshtekarKrishnan}. An 
apparent horizon which is everywhere spacelike coincides 
with a dynamical horizon, but an apparent horizon 
is not required to be spacelike.  Being 
spacelike, dynamical horizons can be crossed only in one 
direction by causal curves, while this is not  
the case for apparent horizons which can be partially or 
entirely timelike.

Finally, {\em slowly evolving horizons} have also been 
introduced and studied \cite{BoothFairhurst04, 
KavanaghBooth06, BoothFairhurst07, 
Boothreview}: these are ``almost 
isolated'' FOTHs and they are intended to represent black hole 
horizons which evolve slowly in time, as is expected in 
many astrophysical processes but not, for example,  in the final 
stages of black  hole evaporation. They are analogous to 
thermodynamical systems in quasi-equilibrium.

\subsection{Kodama vector}

In the literature one finds several notions of surface gravity associated with 
horizons. In stationary situations, in which  a timelike Killing 
vector field  outside the horizon becomes 
null on it, these notions of surface gravity 
coincide. In dynamical situations there is no 
timelike Killing vector and these  surface 
gravities turn out to be 
inequivalent.  In spherical symmetry, the 
Kodama vector mimics the properties of a Killing vector and 
originates  a (miracolously) conserved current and a surface 
gravity.

The Kodama vector \cite{Kodama} is a generalization of the notion 
of Killing vector field to spacetimes which do not admit one, and 
has 
been used in place of  a Killing vector in the thermodynamics of 
dynamically evolving horizons. The Kodama vector is 
defined only for spherically symmetric 
spacetimes (see \cite{Tung08} for an attempt to introduce a Kodama-like 
vector in non-spherical spacetimes). Let the metric be 
\be \label{hab}
ds^2=h_{ab}dx^a dx^b +R^2 d\Omega_{(2)}^2 \,,
\ee
where $\left( a,b \right)=\left(t,R\right)$, $R$ is the areal radius, and 
$d\Omega_{(2)}^2=d\theta^2 +\sin^2\theta \, d\varphi^2 $ is the line
element on the unit 2-sphere. Let 
$\epsilon_{ab}$ 
be the volume form associated with the 2-metric $h_{ab}$ 
\cite{Waldbook}; then the {\em Kodama vector} is \cite{Kodama}
\be
K^a \equiv - \epsilon^{ab} \nabla_b R 
\ee
(with $K^{\theta}=K^{\varphi}=0$). The Kodama vector satisfies
 $ K^a \nabla_a R=- 
\epsilon^{ab} \nabla_a R \nabla_b R =0$.  In a static spacetime the 
Kodama vector is parallel (in general, not equal) to the 
timelike Killing vector. In the region in which it is 
timelike, the Kodama vector defines a class of preferred 
observers with four-velocity $ u^a  \equiv K^a/ \sqrt{ | 
K^c K_c | } $ (the Kodama vector is timelike in asymptotically 
flat regions).

The Kodama vector is divergence-free \cite{Kodama, AbreuVisser10},
\be \label{Kodamadivergence}
\nabla_a K^a=0 \,,
\ee
which  has the consequence that the Kodama energy 
current
\be
J^a \equiv G^{ab}K_b 
\ee
(where $G_{ab}$ is the Einstein tensor) is covariantly 
conserved, $\nabla^aJ_a=0$, \cite{Kodama} even if there is no 
timelike Killing vector (a property referred to as 
the ``Kodama miracle''  \cite{AbreuVisser10}).  By writing 
the spherical metric in  
Schwarzschild-like coordinates, 
\be\label{ABgauge}
ds^2=-A \left(t,R \right) dt^2+B\left(t,R \right) dR^2+R^2 
d\Omega_{(2)}^2 \,,
\ee
the Kodama vector assumes the simple form ({\em e.g.}, \cite{Kodama, Racz06})
\be \label{KodamaAB}
K^a=\frac{-1}{\sqrt{AB}} \left( \frac{\partial}{\partial 
t} \right)^a \,. 
\ee
The Noether charge associated with the Kodama conserved current is the 
Misner-Sharp-Hernandez energy \cite{MisnerSharp, MisnerHernandez} 
of spacetime (which, again, is defined only for 
spherically symmetric spacetimes) \cite{Hayward96PRD53}.

\subsection{Surface gravities}

Traditionally, surface gravity is defined in terms of geometric properties 
of the metric tensor and it also shows up in black hole 
thermodynamics as the proportionality factor between the variation of 
the black hole mass (which plays the role of internal energy) $dM$ and 
the variation of the event horizon area (proportional to the entropy) 
$dA$. Since it is unclear which definition of black 
hole mass is appropriate in non-trivial backgrounds 
(see the review \cite{Szabados04}), also the definition of surface gravity suffers
from the same ambiguities. Surface gravity is also a semiclassical quantity 
since for stationary black holes it coincides, up to a constant, with the Hawking 
temperature of a black hole.

The textbook definition of surface gravity is given on a  
Killing horizon \cite{Waldbook}. Given that Killing fields are not available 
in  non-stationary situations, a different concept of surface gravity is 
necessary there. The recurrent definitions are reviewed 
in Ref.~\cite{Nielsen:2007ac} and are briefly recalled 
here.\\

\noindent {\em Killing horizon surface gravity:} a 
Killing horizon defines the surface gravity $\kappa_{{\small {\sf 
Killing}}}$ as follows \cite{Waldbook}:  on 
the Killing horizon the Killing vector $k^a$ satisfies ({\em e.g.}, 
\cite{Nielsen:2007ac}) 
\be\label{Killingsurfacegravity}
k^a\nabla_a k^b \equiv  \kappa_{{\small {\sf Killing}}} 
\, k^b \,, 
\ee
so $\kappa_{ {\sf Killing}}$ measures the failure of the 
geodesic Killing vector $k^a$ to be affinely parametrized 
on the Killing horizon.  Another property of the Killing surface gravity is 
\cite{Waldbook}
\be \label{Killingsurfacegravity2}
\kappa_{{\small {\sf Killing}}}^2=-\frac{1}{2} \left( 
\nabla^a k^b 
\right) \left( \nabla_a k_b \right) \,. 
\ee
In static spacetimes, $\kappa_{{\small {\sf Killing}}}$ is interpreted 
 as the limiting force required at 
spatial infinity to hold in place a unit test mass just above the 
event horizon  by means of an infinitely long massless string 
\cite{Waldbook} (which shows the non-local 
nature of the notion of Killing surface gravity).  
Since the Killing equation $ \nabla_a k_b + \nabla_b 
k_a=0 $ determines the 
Killing vector $k^a$ only up to an overall normalization, there 
is freedom to rescale $k^a$ and the value of the  
surface gravity depends on the non-affine 
parametrization chosen for $k^a$. However, in static/stationary 
situations one has the freedom of imposing that $k^c k_c=-1$ at 
spatial infinity. The Killing surface gravity can be generalized to any event 
horizon that is not a Killing horizon by replacing the Killing 
vector $k^a$ with the null generator of the event horizon 
\cite{Nielsen:2007ac}.\\

\noindent {\em Surface gravity for marginally 
trapped surfaces:} let $l^a$ and $n^a$ be the  outgoing 
and ingoing null normals to 
a marginally trapped (spacelike compact 2-dimensional) surface, 
with the expansion of $l^a$ vanishing, and assume that $l^a$ and 
$n^a$ are normalized so that $l^c n_c=-1$. In general, $l^a$ is 
not a horizon generator but is a non-affinely parametrized 
geodesic vector on the trapping horizon, which allows one to 
define a surface gravity $\kappa$ as
\be
l^a\nabla_a \, l^b \equiv \kappa \, l^b \,,
\ee
or
\be
\kappa = -n^b l^a \nabla_a \, l_b \,.
\ee
The value of $\kappa$ depends on the parametrization of $l^a$  
and there are various proposals for this. 
In general, writing $l^a$ as the tangent to a null curve 
$x^{\mu}(\lambda) $ with parameter $\lambda$, a  parameter change 
(dependent on the spacetime point)  $\lambda \rightarrow \lambda 
'$ means that the components of $l^a$ change according to
\be
l^{\mu} = \frac{dx^{\mu}}{ d\lambda} \longrightarrow l^{\mu '}
=\frac{ dx^{\mu}}{ d\lambda'} = l^{\mu} \, 
\frac{d\lambda}{d\lambda'} 
\equiv \Omega(x) \, l^{\mu} 
\ee
and 
\begin{eqnarray}
&& l^{\nu '} \nabla_{\nu '} l^{\mu '} = \kappa ' l^{\mu '} 
\,, \\
&&\nonumber\\
&& \Omega l^{\nu}\nabla_{\nu} \left( \Omega l^{\mu} 
\right)=\kappa' 
\Omega \, l^{\mu} \,, 
\end{eqnarray}
and finally
\be
  \kappa \rightarrow \kappa ' = \Omega  \, \kappa 
+l^{c}\nabla_c \Omega \,.
\ee

\noindent The {\em Hayward proposal (for spherical symmetry)} 
\cite{Hayward98CQG} is based 
on the Kodama vector  $ K^a$ \cite{Kodama}. In spherical 
symmetry the Kodama vector  satisfies 
\begin{eqnarray}
&& \nabla_b  \left( K_a T^{ab} \right) \propto \nabla_b J^b =0 
\,, \\
&& \nonumber\\
&& K^c K_c=-1 \;\;\;\;\mbox{ at spatial infinity},
\end{eqnarray}
and it is taken to be future-directed.
The ensuing surface gravity for a trapping horizon is given by
\be \label{xvbcquesta}
 \frac{1}{2}  \, g^{ab} K^c \left( \nabla_c K_a -\nabla_a 
K_c \right) =\kappa_{ {\sf Kodama}} K^b \,. 
 \ee
This definition is unique because the Kodama vector is 
unique. $\kappa_{{\small {\sf Kodama}}}$  agrees with 
the surface gravity on the horizon of a 
Reissner-Nordstr\"om black hole but not with other 
definitions  of dynamical surface gravity.  An expression 
equivalent to (\ref{xvbcquesta}) is \cite{Hayward98CQG}
\be \label{kappaKodamaHayward}
\kappa_{ {\small {\sf Kodama} } } =\frac{1}{2} \, 
\Box_{(h)}R=\frac{1}{2\sqrt{-h}}  \, \partial_{\mu} 
\left( \sqrt{-h} 
\, h^{\mu\nu} \partial_{\nu}R \right) \,,
\ee
where $R$ is the areal radius and  
$h$ is the determinant of the metric $h_{ab}$ in the 
2-space orthogonal to $\nabla_a R$. 
The Hamilton-Jacobi approach, a variant of the Parikh-Wilczek 
method \cite{ParikhWilczek00}, employs the 
Kodama-Hayward definition of surface gravity 
\cite{DiCriscienzoHaywardNadaliniVanzoZerbini10} (for a review 
of tunneling methods see Ref.~\cite{VanzoAcquavivaDiCriscienzo11}).\\

\noindent {\em The Fodor et al. surface gravity:} this proposal 
for spherically symmetric asymptotically flat spacetimes 
\cite{Fodoretal} is based on the ingoing null normal $n^a$ 
being normalized so that $n^a t_a=-1$, where $t^a$ is the 
asymptotic time-translational Killing vector at spatial infinity.
$n^a$ is affinely parametrized everywhere and at spatial infinity 
is parametrized by the proper time of static observers.  
Requiring that $l^c n_c=-1$ fixes the 
parametrization of $l^a$, yielding 
\be
\kappa_{{\small {\sf Fodor}}}= -n^b l^a \nabla_a l_b \,.
\ee

\noindent{\em The  isolated horizon surface gravity:} this proposal of  Ashtekar, 
Beetle, and Fairhurst \cite{AshtekarBeetleFairhurst00} applies to 
an isolated horizon.  $n^a$ is normalized so that its expansion 
agrees with that of the 
Reissner-Nordstr\"om case and with $l^an_a=-1$. This choice fixes 
a unique surface gravity as a function of the horizon parameters. 
However, this concept  appears to be limited, for example 
it cannot be extended to the Einstein-Yang-Mills case 
\cite{AshtekarFairhurstKrishnan00, Nielsen:2007ac}.\\\\

\noindent {\em The Booth and Fairhurst proposal for slowly 
evolving horizons:} this definition \cite{BoothFairhurst04} extends the previous 
proposal. On the 
isolated horizon the normal is $\tau^a=Bl^a +C n^a$, with $B$ and 
$C$ scalar fields defined there, which weight the contributions 
of $l^a$ and $n^a$ (for an isolated horizon it is $B=1 
\,,  C=0$). The surface gravity is
\be
\kappa_{{\sf BF}} \equiv  -B n^a l^b \nabla_b l_a -C l^a 
n^b 
\nabla_b n_a \,.
\ee

\noindent {\em Other proposals:} other proposals for surface gravity include  
Hayward's trapping gravity  \cite{Hayward:1993wb}
\be
\kappa_{{\small {\sf trapping}}} \equiv \frac{1}{2} 
\sqrt{ -n^a \nabla_a \theta_l} 
\ee
and the Mukohyama and Hayward proposal \cite{MukohyamaHayward00}.

The surface gravities listed here are computed in 
Ref.~\cite{Nielsen:2007ac} for a general 
spherically  symmetric metric in Eddington-Finkelstein 
coordinates and in terms of the Misner-Sharp-Hernandez 
mass \cite{MisnerSharp, MisnerHernandez}. Ref.~\cite{PielahnKunstatterNielsen11}
compares these definitions for spherical black holes in 
Painlev\'e-Gullstrand coordinates.

\subsection{Spherical symmetry}

Assuming spherical symmetry greatly simplifies the solution of the 
field equations and the study of horizons. Although this is 
an unrealistic  assumption for rotating astrophysical black holes 
and for  universes with realistic inhomogeneities, it is important for  
the fundamental theory.

In spherically symmetric spacetimes a useful tool is the 
Misner-Sharp-Hernandez mass $M $ 
\cite{MisnerSharp, MisnerHernandez}, which here coincides 
with the Hawking-Hayward quasi-local mass 
\cite{Hawking68, Hayward94}. 
The Misner-Sharp-Hernandez mass is 
defined in GR and for spherical symmetry. Using the areal 
radius $R$ and angular coordinates $\left( \theta, 
\varphi \right)$, a spherical line element can be written as
\be\label{2normal}
ds^2=h_{ab}dx^a dx^b +R^2 d\Omega_{(2)}^2  \;\;\;\;\;\;\; (a,b=1,2).
\ee 
The Misner-Sharp-Hernandez mass $M$ is defined by 
\cite{MisnerSharp, MisnerHernandez}
\be \label{MisnerSharpmass}
1-\frac{2M}{R} \equiv \nabla^c R  \, \nabla_c R 
\ee
or 
\be 
M=\frac{R}{2} \left( 1-h^{ab}\nabla_a R \, \nabla_b R 
\right) \,. \label{MSHz2}
\ee
Horizons in spherical symmetry are discussed in a clear way in the formalism of 
 Nielsen and  Visser \cite{NielsenVisser06, NielsenYeom09}. 
These authors consider the most general spherically symmetric metric 
(not necessarily stationary or asymptotically flat)    
with a spherically symmetric spacetime slicing, which assumes the 
form 
\be \label{generalsphericalmetric}
ds^2=-\mbox{e}^{-2\phi (t, R)} \left[ 1-\frac{2M(t,R)}{R} 
\right] dt^2 +\frac{dR^2}{1-\frac{2M(t,R)}{R} } 
+R^2d\Omega_{(2)}^2 
\ee
in Schwarzschild-like coordinates, where $M(t,R)$ {\em a 
posteriori} turns out to be the  Misner-Sharp-Hernandez 
mass.  The line element (\ref{generalsphericalmetric}) can be 
recast  in Painlev\'e-Gullstrand coordinates as 
\be \label{generalsphericalPG}
ds^2=- \frac{ \mbox{e}^{-2\phi } }{ \left( \partial\tau 
/\partial t\right)^2} \left( 1-\frac{2M}{R} \right) 
d\tau^2 +  \frac{ 2\mbox{e}^{-\phi}}{ \partial\tau/\partial t } 
\sqrt{ \frac{2M}{R}} \, d\tau dR 
+dR^2   +R^2d\Omega_{(2)}^2  \,,
\ee
where $\phi( \tau,R)$ and $M(\tau, R)$ are implicit 
functions of 
$\left( \tau, R \right)$ and the spacelike 
 hypersurfaces $\tau=$constant are flat. Using the 
implicit functions of $ \left( \tau, R \right)$ 
\cite{NielsenVisser06}
\begin{eqnarray}
c \left(\tau, R \right) & \equiv & \frac{ \mbox{e}^{-\phi(t,R)} 
}{\left( \partial\tau/ \partial t \right)}  \,,\\
&&\nonumber\\
v \left(\tau, R \right) & \equiv & 
\sqrt{\frac{2M(t,R)}{R} } \, \frac{ \mbox{e}^{-\phi(t,R)} 
}{\partial\tau/\partial t} =c \,  
\sqrt{\frac{2M}{R} }   \,,
\end{eqnarray}
the line element becomes 
\be\label{vvvucci}
ds^2= -\left[ c^2\left(\tau, R \right) - v^2\left(\tau, R \right) 
\right] d\tau^2 +2v \left(\tau, R \right) d\tau dR +dR^2 +R^2 
d\Omega_{(2)}^2 \,.
\ee
A number of practical results are then obtained \cite{NielsenVisser06}.

The outgoing radial null geodesic congruence has tangent field 
with components (in Painlev\'e-Gullstrand coordinates $\left( 
\tau, R, \theta, \varphi \right)$) 
\be
l^{\mu}=\frac{1}{c(\tau, R)} \Bigg( 1, c(\tau, R)-v(\tau, R), 0,0 
\Bigg) \,,
\ee
while the ingoing radial null geodesics have tangent field 
\be
n^{\mu}=\frac{1}{c(\tau, R)} \Bigg( 1, -c(\tau, R)-v(\tau, R), 
0,0 \Bigg) \,,
\ee
where the normalization 
\be
g_{ab}l^a n^b=-2
\ee
is adopted \cite{NielsenVisser06}.  The expansions of these 
radial null geodesic congruences are
\begin{eqnarray}
\theta_l &=& \frac{2}{R} \left( 1-\sqrt{ \frac{2M}{R} } \right) 
\,,\\
&&\nonumber\\
\theta_n &=& - \, \frac{2}{R} \left( 1 +\sqrt{ \frac{2M}{R} 
} \right) 
\,. \label{2.159}
\end{eqnarray}
A  sphere of radius $R$ is 
\cite{Hayward96PRD53, Poissonbook, NielsenVisser06}
trapped if $R<2M $, marginal if $R=2M $, and  
untrapped if $R>2M $. The apparent horizon 
 corresponding to $\theta_l=0$ and 
$\theta_n<0$ is given  by 
\be \label{AH}
\frac{2M\left( \tau, R_{AH}\right)}{R_{AH}(\tau)}=1
 \;\; \Longleftrightarrow \;\; \nabla^c 
R\nabla_c R \left. \right|_{AH} =0 \Longleftrightarrow \;\; 
g^{RR} \left. \right|_{AH}=0 \,,   
\ee
where the last equation holds in both Painlev\'e-Gullstrand 
coordinates and in the gauge (\ref{generalsphericalmetric}) and is obtained 
by using the fact that the inverse of the metric (\ref{vvvucci}) has components
\be
\left( g^{\mu\nu} \right)=\frac{1}{c^2} \left(
\begin{array}{cccc}
1 & -v &0 &0 \\
&&&\\
-v & -(c^2-v^2) &0 &0 \\
&&&\\
0 & 0 & \frac{1}{R^2} &0 \\
&&&\\
0 & 0 & 0 &  \frac{1}{R^2\sin^2 \theta} \\ 
\end{array} \right) \,.
\ee
In practice, the condition $g^{RR}=0$ is a very convenient recipe 
to locate the apparent horizons in the presence of 
spherical symmetry when the areal radius $R$ is used as a coordinate 
and it is often convenient to perform a coordinate transformation 
to this radial coordinate and to rewrite the line element using $R$.

The gradient of the 
areal radius $R$ and the normal 
$n_a=\nabla_a R$ to the surfaces $R=$const. become null at the 
apparent  horizon; this recipe is reminiscent of the change in 
the causal 
character of the Schwarzschild radial coordinate
 on the Schwarzschild event horizon. 
However the apparent horizon is not, in general, a null surface. We have also 
\cite{NielsenVisser06}
\be
{\cal L}_{n} \theta_l \left. \right|_{AH}=n^a \nabla_a \left[ 
\frac{2}{R}  \left( 1-\sqrt{ \frac{2M}{R}} \right) 
\right]_{AH}
=- \frac{2 \left(1-2M'_{AH} \right)}{R_{AH}^2} \left( 
1+\frac{\dot{R}_{AH} }{2c_{AH}} \right) \,,
\ee
where a prime and an overdot denote partial differentiation with 
respect to $R$ and $\tau$, respectively, and the subscript $AH$  
identifies quantities evaluated on the apparent horizon. 
$1-2M'_{AH}>0$ is required for the horizon to be outer in a 
spacetime with regular asymptotic region, hence the 
condition for the apparent horizon to be also a trapping horizon 
is \cite{NielsenVisser06, NielsenYeom09} 
\be
 \dot{R}_{AH} >-2c_{AH} \,.
\ee 
If matter satisfies the null energy condition, and assuming the 
Einstein equations, the area of the apparent horizon cannot 
decrease. Various energy fluxes across the apparent horizon are 
also discussed and computed  in \cite{NielsenVisser06}. The 
Nielsen-Visser  surface gravity at the horizon is computed 
from $ l^b \nabla_b 
l^a=\kappa_l \, l^a $, which gives \cite{NielsenVisser06}
\be
\kappa_l ( \tau ) = \frac{ 1-2M' \left( \tau, R_H( \tau) 
\right) }{ 2 R_H( \tau) }  \,.
\ee 
An extremal horizon will be one with vanishing 
surface gravity, 
\be 
1-2M'\left( \tau, R_H(\tau) \right)=0 \,.
\ee

The fact that the Misner-Sharp-Hernandez mass can be used to define and locate 
apparent horizons in spherically symmetric spacetimes shows that the apparent horizon 
is a quasi-local concept and is independent of the global causal structure. However, 
it does not appear to be a completely local notion (it depends on a surface, not only 
on the spacetime point).

\section{Evolving horizons, cosmological black holes, and naked singularities in GR}

Let us review briefly some dynamical and spherically symmetric 
solutions of the Einstein equations of particular 
significance, paying attention to the structure and dynamics 
of their apparent horizons.

\subsection{The Schwarzschild-de Sitter-Kottler spacetime}

The Schwarzschild-de Sitter-Kottler \cite{Kottler18} spacetime is locally 
static but it is useful to review it in order to understand the apparent 
horizons of more complicated dynamical solutions. It has line element 
\be 
ds^2=-\left( 
1-\frac{2m}{R}-H^2R^2 \right) dt^2+ \left( 1-\frac{2m}{R}-H^2R^2 \right)^{-1} 
dR^2+ R^2 d \Omega_{(2)}^2 \,,\label{staticSdS} 
\ee 
where the constant 
$H=\sqrt{\Lambda/3} $ is the Hubble parameter of the de Sitter background, 
$\Lambda>0$ is the cosmological constant, and $m>0$ is a second parameter 
related to the mass of the central inhomogeneity ({\em e.g.}, 
\cite{HawkingEllis73, Boussohepth0205177}). The static coordinates $\left( t, R 
,\theta,\varphi \right)$ cover the region $ R_1 <R <R_2 $. The apparent 
horizons are located by $g^{RR}=0$, which is equivalent to the cubic equation 
\be\label{3} 
1-\frac{2m}{R} -H^2R^2 =0 \,, 
\ee 
with roots 
\begin{eqnarray} 
R_1&=&\frac{2}{\sqrt{3}H}\sin\psi \,,\nonumber\\ 
&&\nonumber\\ 
R_2&=&\frac{1}{H}\cos\psi -\frac{1}{\sqrt{3}H}\sin\psi \,,\nonumber\\ 
&&\nonumber\\ 
R_3&=&-\frac{1}{H}\cos\psi -\frac{1}{\sqrt{3}H} \sin\psi \,, 
\end{eqnarray} 
with $\sin (3\psi )=3\sqrt{3} \, mH$.  $m$ and $H$ are both 
necessarily positive in an expanding universe, then $R_3$ is negative and there 
are at most two apparent horizons. When $R_1$ and $R_2$ are real, $R_1$ is a 
black hole apparent horizon which reduces to the $R=2m$ Schwarzschild horizon 
in the limit $H\rightarrow 0$, while $R_2$ is a cosmological apparent horizon 
which reduces to the $R=1/H$ de Sitter horizon in the limit $m \rightarrow 0$. 
The metric (\ref{staticSdS}) is static in the region between these two 
horizons.

Both apparent horizons exist only if 
$ 0<\sin( 3\psi ) < 1$ and, since the metric is 
locally static, the apparent black hole 
and  cosmological horizons are also event horizons. If 
$\sin (3\psi)=1$  these horizons 
coincide (extremal Nariai black hole).  For $\sin 
(3\psi ) >1$ the roots are complex-valued and there is  
a naked singularity. To summarize:
if $ mH<1/(3\sqrt{3})  $ there are two  
horizons of radii $R_1 $  
and $R_2$; if $ mH=1/(3\sqrt{3}) $ the two horizons 
coincide, $R_1=R_2$; if $ mH>1/(3\sqrt{3})$ there are no apparent 
horizons. The  interpretation seems to be that the would-be 
 black hole horizon would become larger than the 
cosmological one but, strictly speaking, the roots corresponding to the 
apparent horizons  are complex in this case.

The black hole horizon has area $ {\cal A} = 4 \pi R_1^2$ which is, of course, 
time-independent. In the non-extremal case the central singularity is eternal and 
spacelike (\cite{GriffithsPodolskyBook}, see this reference also for a conformal 
diagram) and is surrounded by the black hole event horizon at all times for the 
parameter values for which this horizon exists.

A sphere of radius $R$ has the Misner-Sharp-Hernandez mass 
\be
M_{MSH}=m+\frac{H^2R^3}{2} =m+\frac{4\pi}{3} \, \rho R^3\,,
\ee
where $\rho=\frac{\Lambda}{8\pi} $. The Schwarzschild-de Sitter-Kottler 
black hole has been studied extensively in relation to its  
thermodynamics. 
 Here we do not discuss anti-de Sitter black holes corresponding to 
$\Lambda<0$, which are the subject of much recent interest due to the
 fluid-gravity duality  \cite{FiguerasHubenyRangamaniRoss09, 
Hubeny11}.

\subsection{The McVittie solution}

The 1933 McVittie solution of the Einstein equations \cite{McVittie} is 
a generalization of the Schwarzschild-de Sitter-Kottler solution and 
represents a central object embedded in a FLRW (not necessarily a 
locally static de Sitter) background. Even after many works 
\cite{Sussman85, Krasinskibook, NolanPRD98, Nolanb, Nolan2, Kleban, 
Roshina1, Roshina2, AndresRoshina, LakeAbdelqader11, Anderson11, 
SilvaFontaniniGuariento12}, this solution is not completely understood. 
The McVittie solution with negative cosmological constant was analyzed 
in Ref.~\cite{LandryAbdelqaderLake12} and an electrically charged 
version of the McVittie spacetime was found in Ref.~\cite{GaoZhang06}. 
In this subsection we restrict to a spatially flat FLRW background  
and to zero electric charge.
 
A simplifying assumption of McVittie 
consists of the no-accretion condition $G_t^{\bar{r}}=0$ (in 
spherical coordinates, where 
$G_{\mu\nu}$ is the Einstein tensor) which forbids any 
mass-energy flow (which, in spherical symmetry, could only be radial), 
${T_t}^{\bar{r}}=0$. 
Generalizations of the McVittie 
solution allowing radial energy fluxes are more complicated 
and will be considered later. McVittie was led to his 
solution   
\cite{McVittie} by the problem of  the 
effect of the cosmological 
expansion on local systems. Different approaches  
to this  problem generated other solutions, such as 
the  Swiss-cheese model    
\cite{EinsteinStraus, EinsteinStraus2} (this problem has seen   
an extensive literature devoted to it but is  not  
completely solved \cite{CarreraGiuliniRMD10}). Unlike the 
Schwarzschild-de Sitter-Kottler spacetime, black 
holes in more general FLRW  backgrounds are 
dynamical.

The McVittie line element in isotropic coordinates is
\begin{equation} \label{McVittieisotropic}
ds^2=-\frac{  \left(1-\frac{m(t)}{2\bar{r}} \right)^2}{
\left(1+\frac{m(t)}{2\bar{r}} \right)^2} \, dt^2+
a^2(t) \left( 1+\frac{m(t)}{2\bar{r}} \right)^4 \left( 
d\bar{r}^2 +\bar{r}^2 d\Omega_{(2)}^2 \right) \,,
\end{equation}
where the function $m(t)$ is required to  satisfy  
the McVittie 
no-accretion condition $T_t^{\bar{r}}=0$ 
 on the stress-energy tensor $T_{ab}$, 
which becomes
\begin{equation}  \label{35}
\frac{\dot{m}}{m}+\frac{\dot{a}}{a}=0 
\end{equation}
with solution 
\begin{equation} \label{36}
m(t)=\frac{m_0}{a(t)} \,,
\end{equation}
where $m_0 $ is a constant, therefore, 
\begin{equation} 
ds^2=-\frac{  \left[ 1-\frac{m_0}{2\bar{r}a(t)} 
\right]^2}{
\left[ 1+\frac{m_0}{2\bar{r}a(t)} \right]^2} \, dt^2+
a^2(t) \left[ 1+\frac{m_0}{2\bar{r}a(t)} \right]^4 \left( 
d\bar{r}^2 +\bar{r}^2 d\Omega_{(2)}^2 \right) \,.
\end{equation}
The McVittie metric reduces to the Schwarzschild one  
in isotropic  coordinates if $ a \equiv 1$ and to the 
FLRW metric if $m_0 = 0$ and is 
singular on the 
2-sphere $\bar{r}=m_0/2$ (which reduces to the 
Schwarzschild horizon if $a \equiv 1$)  \cite{Ferrarisetal, 
NolanPRD98, Nolanb, Nolan2, Sussman85}. This singularity 
is spacelike 
\cite{NolanPRD98, Nolanb, Nolan2, 
LakeAbdelqader11, SilvaFontaniniGuariento12} 
(and is represented as a horizontal line
in conformal diagrams). There is another 
spacetime singularity
at $\bar{r}=0$.  McVittie's original interpretation 
of the line element (\ref{McVittieisotropic}) as  
describing a point mass  at  $\bar{r}=0$ is made untenable 
by the fact that this 
point mass would be  surrounded by the  
$\bar{r}=m_0/2$ singularity
\cite{Sussman85, Ferrarisetal, NolanPRD98, Nolanb, Nolan2}. 
We will only consider the region $\bar{r}>2m_0$ here. 
The energy density of the 
source fluid is finite but its  pressure 
\begin{equation} \label{pressure}
P \left( t, \bar{r} \right) 
=-\, \frac{1}{8\pi} \left[ 3H^2+\frac{2\dot{H}\left( 
1+\frac{m_0}{2\bar{r}} \right)  }{1-\frac{m_0}{2\bar{r}} }\right]
\end{equation}
diverges at $\bar{r}=m_0/2$  with  the 
Ricci scalar ${R^a}_{a}=8\pi \left( 3P-\rho \right)$
\cite{Sussman85,Ferrarisetal, NolanPRD98, Nolanb, Nolan2, McClureDyer06CQG, 
McClureDyerGRG}, with 
the exception of a de Sitter background with
 $\dot{H}=0$  \cite{NolanPRD98, 
Nolanb, Nolan2, AudreyPRD, Anderson11}.

The  apparent horizons were studied in Refs.~\cite{Nolanb, LiWang06, AndresRoshina} and 
interpreted in \cite{AndresRoshina}, which we follow here. We rewrite 
the  line element (\ref{McVittieisotropic}) 
in terms of the areal radius
\be
R\left( t, \bar{r} \right) 
\equiv a(t) \bar{r} \left( 1+\frac{m}{2\bar{r}} \right)^2 
\,;
\ee
the differentials $d\bar{r}$ and $dR$ are related by 
\begin{eqnarray}
dR &=& \left( 1+\frac{m}{2\bar{r}} \right) a\bar{r} 
\left[ H \left( 1+\frac{m}{2\bar{r}} \right) 
+\frac{\dot{m}}{\bar{r}} \right] dt 
 + a  \left( 1+\frac{m}{2\bar{r}} \right)  
\left( 1 - \frac{m}{2\bar{r}} \right) d\bar{r} \nonumber\\
&&\nonumber\\
&=& a \left( 1+\frac{m}{2\bar{r}}\right)
\left( 1-\frac{m}{2\bar{r}}\right) 
\left( H\bar{r}dt + d\bar{r} \right) \,,
\end{eqnarray}
where the relation (\ref{35}), which  gives 
\be
H \left( 1+\frac{m}{2\bar{r}} \right) 
+\frac{\dot{m}}{\bar{r}} = 
H \left( 1 - \frac{m}{2\bar{r}} \right) \,,
\ee
has been used and 
\be
d \bar{r}=\frac{dR}{ a\left( 1+\frac{m}{2\bar{r}} \right) 
\left( 1 - \frac{m}{2\bar{r}} \right)} -H\bar{r} dt \,.
\ee
Using this relation in  
(\ref{McVittieisotropic}) and noting that
\be
 \left( \frac{ 1-\frac{m}{2\bar{r}} }{
1+\frac{m}{2\bar{r}} } \right)^2= 1-\frac{2m_0}{R}
\ee
where $ m/\bar{r}=ma/R=m_0/R$ (here $ma$ is constant because 
of eq. (\ref{35})) leads to 
\be \label{nondiagonalMcVittie}
ds^2 = -\left( 1-\frac{2m_0}{R}-H^2R^2 \right) dt^2
+\frac{dR^2 }{1-\frac{2m_0}{R} } 
-\frac{2HR}{\sqrt{1-\frac{2m_0}{R} } } \, dtdR +R^2 
d\Omega_{(2)}^2 \,,
\ee
where $H\equiv \dot{a}/a$. The cross-term in $dtdR$ is then  
eliminated by defining a new time  $T( t, R)$ 
such that 
\be \label{formdT}
dT=\frac{1}{F} \left( dt+\beta dR \right) \,,
\ee
with the integrating factor $F(t, R)$ and function 
$\beta(t,R)$ to be determined.  $dT$ 
is an exact differential if the 1-form (\ref{formdT}) is  
closed, or 
\be
\frac{\partial F}{\partial R}=
F\, \frac{\partial \beta}{\partial t} 
-\beta \, \frac{\partial F}{\partial t} \,.
\ee
Now replace $dt$ with $FdT-\beta dR$ in eq. 
(\ref{nondiagonalMcVittie}), obtaining
\begin{eqnarray}
ds^2 & = & -\left( 1-\frac{2m_0}{R}-H^2R^2 \right) F^2 
dT^2\nonumber\\
&&\nonumber\\
&\, & + \left[ -\left( 1-\frac{2m_0}{R} -H^2R^2 
\right)\beta^2 + \frac{1}{ 1-\frac{2m_0}{R} } 
 +\frac{2\beta HR}{ \sqrt{ 
1-\frac{2m_0}{R} }} \right] dR^2 \nonumber\\
&&\nonumber\\ 
&\,& + 2F \left[ \left( 1-\frac{2m_0}{R} -H^2R^2 \right) 
\beta  - \frac{HR}{ \sqrt{1-\frac{2m_0}{R}}} \right] dT 
dR +R^2 d\Omega_{(2)}^2 \,.\nonumber\\
&&
\end{eqnarray}
Imposing now that 
\be
\beta (t, R) = 
\frac{HR}{  \sqrt{1-\frac{2m_0}{R}}\, 
  \left( 1-\frac{2m_0}{R}-H^2R^2 \right) } \,,
\ee
the line element is diagonalized,
\be \label{diagonalMcVittie}
ds^2 = -\left( 1-\frac{2m_0}{R}-H^2R^2 \right) F^2 
dT^2
+\frac{dR^2 }{1-\frac{2m_0 }{R}-H^2R^2  } 
+R^2  d\Omega_{(2)}^2 \,. 
\ee
The singularity $\bar{r}=m/2$ corresponds to the 
proper radius $R=2m \, a(t)=2  m_0$ 
and does not expand.

Let us study now the apparent horizons of the McVittie spacetime 
\cite{NolanPRD98, Nolanb, AndresRoshina}. 
For simplicity, we restrict ourselves 
to a spatially 
flat FLRW background. The Einstein equations provide  
the density of the  fluid 
\be
\rho(t)=\frac{3}{8\pi} \, H^2(t) \,,
\ee
The McVittie metric admits arbitrary FLRW backgrounds 
generated by cosmic fluids 
satisfying any constant equation of state. For brevity,  we restrict  
to a cosmic fluid which reduces 
to dust at spatial infinity and corresponding to an
equation of state parameter $w=0$. Then the pressure is
 \cite{Roshina1, Roshina2}
\be
 P\left(t,R \right)=\rho(t) \left( 
\frac{1}{\sqrt{1-\frac{2m_0}{R}}} -1 \right) \label{5} \,.
\ee
The apparent horizons are located by   
$g^{RR}=0$ or, using eq. (\ref{diagonalMcVittie}) 
\be\label{8}
1-\frac{2m_0}{R}-H^2(t) \, R^2=0 \,.
\ee
This cubic in $R$ is the same as the  
Schwarzschild-de  Sitter-Kottler horizon condition 
(\ref{3}) but with a time-dependent Hubble 
parameter. The resulting time-dependent apparent 
horizons $R_1(t)$ and $R_2(t)$ are again  
the  solutions $R_{1,2}$ of eq. (\ref{3})  
but now with time-dependent coefficient  $H(t)$. 
The location of the apparent horizons of 
the  McVittie spacetime depends on the cosmic time. Again, the 
condition for both horizons to exist is $0<\sin 
( 3\psi ) <1$, which corresponds to 
$m_0H(t)<1/(3\sqrt{3})$.  
However, unlike the Schwarschild-de Sitter-Kottler case 
with constant $H$, this inequality is only 
satisfied at certain times.  The critical time at which 
$m_0 H(t)=1/(3\sqrt{3})$ is unique for a dust-dominated background with 
$H(t)=2/(3t)$ and is  $t_* =2\sqrt{3} \, m_0$.  Three possibilities arise:

\begin{enumerate}

\item  for  $t<t_*$ it is $m_0>\frac{1}{3\sqrt{3} 
\,H(t)}$ and both $R_1(t)$ and $R_2(t)$ are complex. 
There are no apparent horizons.

\item The critical time $t=t_*$ corresponds to  
$m_0=\frac{1}{3\sqrt{3}\,H(t)}$.  $R_1(t)$ and 
$R_2(t)$ coincide at a real value and there 
is a single apparent horizon at $R_*=\frac{1}{\sqrt{3}\,H(t_*)}$.
 
\item For $t>t_*$ it is $ m_0 < 
\frac{1}{3\sqrt{3}\,H(t)}$ and there are two apparent horizons of 
real positive radii $R_1(t)$ and 
$ R_2(t)$. 

\end{enumerate}

The behaviour of the apparent horizons is described  in 
fig.~\ref{McV1}.
\begin{figure}
\centering
\includegraphics[width=8.5cm]{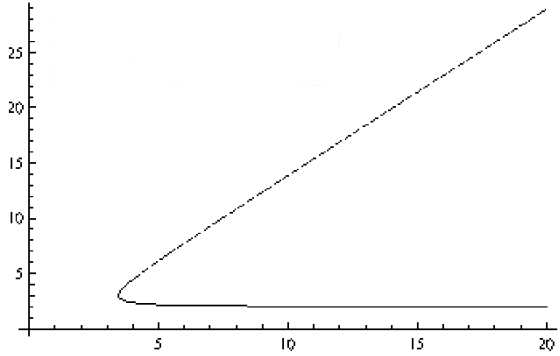}
\caption{The McVittie cosmological (dashed) and black hole 
(solid) apparent  horizons in a dust-dominated background 
universe. Time $t$ (on the horizontal axis) and radius $R$ (on the 
vertical axis) are  in units of $m_0$ and we arbitrarily fix  $m_0=1$.
\label{McV1}}
\end{figure}
 At times  $t<t_{*}$ there is a  
naked singularity at $R=2m_0$: while   
the Hubble parameter $H(t)$ diverges near the Big Bang, 
the mass coefficient $m_0$ stays supercritical at
 $m_0>\frac{1}{3\sqrt{3}\,H(t)}$.  
As the Schwarzschild-de Sitter-Kottler experience teaches us, 
 a black hole horizon cannot 
be accommodated in this small universe and the singularity is 
naked: the putative black hole is too large to fit in the observable universe 
(varying speed of light cosmologies have a related phenomenology --- the radii  
of primordial black holes and the Compton wavelengths of massive particle states 
can become larger than the Hubble radius \cite{BarrowVSL}).
At the critical time $t_{*}$ a black hole apparent horizon 
and a  cosmological apparent 
horizon appear together at radius $R_1(t_*)=R_2(t_*)
=\frac{1}{\sqrt{3}\,H(t_*)}$, in analogy with the Nariai 
black hole of the Schwarzschild-de Sitter-Kottler 
solution. This critical black hole is instantaneous.  
As time progresses to $t>t_{*}$, this single horizon splits 
into an evolving black hole apparent horizon surrounded by 
an evolving cosmological horizon.   The black hole apparent horizon shrinks, 
asymptoting to the spacetime singularity at $2m_0$ 
from above as $t\rightarrow +\infty$, while the 
cosmological apparent horizon expands monotonically, 
tending to $1/H(t)$ in the same limit.

The well known singularity $R=2m_0$ \cite{NolanPRD98, Nolanb, Nolan2, 
Roshina1, Roshina2, LakeAbdelqader11}, where
 the Ricci scalar
\be
{R^a}_a=-8 \pi T^{\mu}_{\mu}=8\pi \left( \rho -3P 
\right)=8\pi 
\rho(t) \left( 4-\frac{3}{\sqrt{1-\frac{2m_0}{R} } } 
\right) 
\ee
diverges, separates the two disconnected spacetime regions $R<2m_0$ 
and $R>2m_0$ \cite{NolanPRD98, Nolanb, Nolan2} and is spacelike 
\cite{AndresRoshina}. One can compare the rate of change of 
the apparent horizon radii  with respect to that of the cosmic substratum, 
obtaining \cite{AndresRoshina}
\be
\frac{\dot{R}_{AH} }{R_{AH} }-H= -H\left( 1+ 
\frac{2\dot{H}R_{AH}^2 }{3H^2 R_{AH}^2 -1} \right) 
\,:\label{eq:rateofexp}
\ee
the apparent horizons are not comoving except for trivial cases. 
The sum of the areas of the two   
apparent horizons of the McVittie spacetime is a 
non-decreasing function of time but undergoes a discontinuous jump from 
zero  at the critical time $t_*$ \cite{AndresRoshina}.

\subsubsection{A phantom background}

A background FLRW universe  
dominated by a phantom fluid with equation of state
parameter $w \equiv P/\rho<-1$ which  violates  the weak 
energy condition can be considered. Phantom fluids,  studied in 
conjunction with  the present cosmic acceleration  
\cite{AmendolaTsujikawabook}, cause  a Big 
Rip singularity at a finite future 
$t_{rip}$ \cite{Caldwell}. A phantom background  
universe for the  McVittie solution was studied in Ref. 
\cite{AndresRoshina}. The scale factor of a  
phantom-dominated spatially flat FLRW universe is 
\be
a(t)=\frac{A}{ \left( t_{rip}-t 
\right)^{\frac{2}{3| w+1|} }}, 
\ee 
where $A$ is a constant. The Hubble parameter 
\be
H(t)=\frac{2}{3|w+1|} \, \frac{1}{t_{rip}-t} 
\ee
is qualitatively the time-reverse of  
that of a dust-dominated universe $H(t)=2/(3t)$.  
  The Hubble parameter for a phantom fluid  
is finite at $t=0$ and increases 
until the Big Rip, at which it  diverges.  The 
apparent horizons around McVittie black holes  embedded in a phantom 
fluid  behave in the opposite 
way to those in a background with $w>-1$ \cite{AndresRoshina} 
(fig.~\ref{McV2}).

\begin{figure}
\centering
\includegraphics[width=8.5cm]{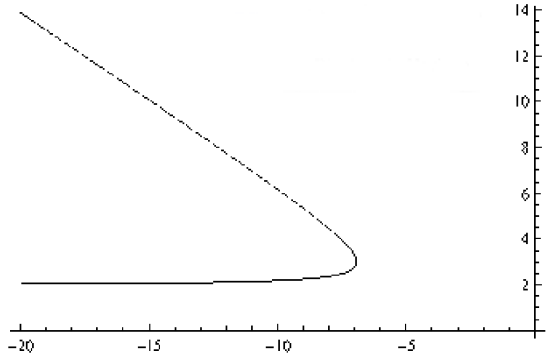}
\caption{\label{McV2} The radii of the McVittie apparent horizons 
(vertical axis) versus time (horizontal axis)  in a 
phantom-dominated universe (here $w=-1.5$ and $t_{rip}=0$).
}
\end{figure}

An idealized interior  solution for the McVittie metric 
describing  a relativistic star of uniform density in a FLRW 
background was found by Nolan \cite{NIS} and it generalizes 
the Schwarzschild interior solution with a Minkowski 
background \cite{Waldbook}, to which it reduces when $a=$const. It 
belongs to  
the Kustaanheimo family of shear-free solutions. 
The star surface is comoving with the cosmic substratum 
\cite{AudreyPRD}. The generalization of the  
Tolman-Oppenheimer-Volkoff equation \cite{Waldbook}   
for this crude star model was written down in~\cite{AudreyPRD}.

Recent works on the McVittie spacetime study its conformal structure 
\cite{Kleban, LakeAbdelqader11, SilvaFontaniniGuariento12}, which means 
integrating numerically the null geodesics or deriving general analytical 
results upon assuming something on the expansion. Lake and Abdelqader 
\cite{LakeAbdelqader11} find that null geodesics asymptote to the 
singularity without entering it. Depending on the form of the scale factor, 
a bifurcation surface may appear which splits the spacetime boundary into a 
black hole horizon in the future and a white hole horizon in the past. This 
behaviour seems a reflection of the McVittie no-accretion condition which 
applies to a timelike dust and, in the limit, also to a null dust. da Silva 
{\em et al.} \cite{SilvaFontaniniGuariento12} find that the presence or 
absence of this white hole horizon depends crucially on the expansion 
history of the universe $a(t)$, and prove a theorem in this regard for 
McVittie spacetimes for which the background is non-superaccelerating ({\em 
i.e.}, $\dot{H} \leq 0$) and de Sitter at late times. It would be desirable 
to extend the result to backgrounds which at late times asymptote to any 
FLRW space, not just de Sitter. See Refs. \cite{LakeAbdelqader11, 
SilvaFontaniniGuariento12} for the corresponding conformal diagrams.

\subsection{Area quantization and McVittie solutions as toy models}

As an example of the use of cosmological black holes as toy models to exemplify 
unintuitive physics, we quote the current issue of the quantization of black 
hole areas. Inspired by certain stringy black holes, there has been 
excitement in the 
string community about the fact that the areas $A_{\pm} $ of black hole inner 
($-$) and outer ($+$) horizons satisfy the relation
\be\label{expression1} 
A_{\pm} = 8\pi l_{pl}^2 
\left( \sqrt{N_1} \pm \sqrt{N_2} \, \right) \,, \;\;\;\;\;\;\;\; N_1, N_2 \in 
\mathbb{N} \,, 
\ee 
or 
\be\label{expression2} 
A_{+} A_{-} =\left( 8\pi 
l_{pl}^2\right)^2 N \,, \;\;\;\;\;\;\;\; N \in \mathbb{N} \,, 
\ee 
where $l_{pl}$ is the Planck length \cite{Larsen97,CveticLarsen97}. 
These area-quantizing relations have somehow come to be seen as universal 
\cite{AnsorgHennig08, AnsorgHennig09, CastroRodriguez12, Castroetal13}. While 
certain stringy black holes remarkably do satisfy these relations, this property 
is certainly not universal, as shown by Visser \cite{VisserBHareas1, 
VisserBHareas2} using 4-dimensional GR black holes. The McVittie solutions 
provide further, and even more convincing examples: if eq.~(\ref{expression1}) 
or eq.~(\ref{expression2}) is satisfied at an instant of time, it fails at 
subsequent times due to the dynamical character of the horizons \cite{BHareas}, 
and realistic black holes are dynamical if nothing else because of Hawking 
radiation and of quantum fluctuations.

\subsection{Generalized McVittie solutions}

Generalized McVittie solutions with spacetime metric of the form  
(\ref{McVittieisotropic}}) but without  the  no-accretion restriction 
(\ref{35}) were introduced in Ref.~\cite{AudreyPRD}. In principle  
such metrics could be meaningless: in the 
``Synge approach'' one can always 
impose that an invented metric solves the 
Einstein equations  and run these equations from  
left to right to compute the corresponding formal stress-energy tensor 
$T_{ab}$. This $T_{ab}$ is 
usually found to be completely unphysical and violates all 
reasonable  energy conditions, beginning with the 
positivity of the energy density. Rather 
surprisingly, generalized McVittie solutions 
with reasonable  matter sources exist.

In isotropic coordinates, generalized McVittie solutions can be 
presented as 
\begin{equation} 
ds^2= -\frac{B^2\left(t, \bar{r} \right)}{
A^2\left(t, \bar{r} \right)}\, dt^2 +a^2(t) A^4 \left(t, \bar{r} 
\right) \left( d\bar{r}^2+\bar{r}^2 d\Omega^2_{(2)} \right) \,,
\end{equation}
where $m(t) \geq 0$ and 
\begin{equation}  
A \left(t, \bar{r} \right) = 1+\frac{m(t)}{2\bar{r}} \,, 
\;\;\;\;\;\;\;
B \left(t, \bar{r} \right) = 1-\frac{m(t)}{2\bar{r}} \,.
\end{equation}
The only non-vanishing components of the mixed Einstein 
tensor are 
\begin{eqnarray}
G_t^t &=& -\, \frac{3A^2}{B^2}\left( 
\frac{\dot{a}}{a} +\frac{\dot{m}}{\bar{r}A} \right)^2 \,, 
\label{einst1} \\
&&\nonumber \\
G_t^{\bar{r}} &=&  \frac{2m}{ \bar{r}^2 a^2 A^5 B} \left( 
\frac{\dot{m}}{m} + \frac{\dot{a}}{a} \right) \,, \\
&&\label{einst2} \nonumber \\
G_{\bar{r}} ^{\bar{r}}  &=& G_{\theta}^{\theta} =G_{\varphi}^{\varphi} =
- \frac{A^2}{B^2}\left\{ 
2  \frac{d}{dt} \left( 
\frac{\dot{a}}{a}+\frac{\dot{m}}{\bar{r}A} 
\right) 
+ \left( \frac{\dot{a}}{a}+\frac{\dot{m}}{\bar{r}A} \right) 
\right.\nonumber \\
&& \nonumber \\
&& \left. \cdot \left[
3 \left( \frac{\dot{a}}{a}+\frac{\dot{m}}{\bar{r}A} \right)
+ \frac{2\dot{m}}{\bar{r}AB} \right]\right\}  
\label{einst3}
\end{eqnarray}
(the unusual feature that $G^{\bar{r}}_{\bar{r}} =G^{\theta}_{\theta}$ 
is named ``spatial Ricci isotropy'' 
in Ref.~\cite{CarreraGiuliniPRD10}).
The quantity
\begin{equation} 
C\equiv  \frac{\dot{a}}{a}+\frac{\dot{m}}{\bar{r}A} =
\frac{\dot{M}}{M}-\frac{\dot{m}}{m} \frac{B}{A}
\end{equation}
appearing in the Einstein tensor reduces to $ 
\dot{M}/M $ where 
\be
M(t) \equiv m(t)a(t)
\ee
for the special subclass of solutions with  $m=$constant. This 
subclass will be called ``comoving mass'' 
solutions.  On the surface 
$\bar{r}=m/2$, $C$  reduces 
to 
\begin{equation}  
C_{\Sigma}=
\frac{\dot{a}}{a}+\frac{\dot{m}}{m} =
\frac{\dot{M}}{M}  
\end{equation}
  for any function $m(t)$.  McVittie solutions correspond to $C_{\Sigma}=0$, while 
comoving mass solutions have $C=C_{\Sigma}=H$ everywhere.

The Ricci scalar  
\begin{equation} 
{R^a}_a = \frac{3A^2}{B^2}\left( 2\dot{C} 
+4C^2 +\frac{ 
2\dot{m}C}{\bar{r}AB} \right)
\end{equation}
diverges on the surface $\bar{r}=m/2$ unless $m$ is 
a constant. Imperfect  fluids can 
be  contemplated as matter sources for this metric.

\subsubsection{Single perfect fluid}

If the matter source of the generalized McVittie metric is 
a single perfect fluid with stress energy tensor 
\begin{equation} 
T_{ab}=\left( P+\rho \right)u_a u_b +Pg_{ab} 
\end{equation}
and a radial fluid flow described by the 
fluid  four-velocity $ u^{\mu}=\left( u^0, u, 0,0 \right) 
$, is allowed, then the only possible solution of the Einstein 
equations is the Schwarzschild-de  Sitter-Kottler black 
hole \cite{AudreyPRD, GaoChenFaraoniShen08, 
LakeAbdelqader11}. This is easily seen, since the normalization 
$u^c u_c=-1 $ yields
\begin{equation} 
u^t=\frac{A}{B} \, \sqrt{ 1+a^2 A^4 u^2} 
\end{equation}
and, using eqs.~(\ref{einst1})-(\ref{einst3}), the 
Einstein equations imply that 
\begin{equation}  \label{delta0}
\dot{M}=- B^2 au \left( P+\rho \right) {\cal A} \sqrt{ 
1+a^2A^4 u^2} \,,
\end{equation}
where  
\be
{\cal A}=\int \int d\theta d\varphi \, 
\sqrt{g_{\Sigma}}=4\pi a^2 A^4 \bar{r}^2
\ee
is the area of a sphere of isotropic  radius $\bar{r}$ and  
\begin{eqnarray} 
&& 3\left( \frac{AC}{B} \right)^2=8\pi  \left[ \left( P+\rho 
\right)a^2A^4 u^2 +\rho \right] \,, \\
&&\nonumber \\
&& -\left( \frac{A}{B} \right)^2 \left( 2\dot{C}+3C^2 
+\frac{2\dot{m}C}{\bar{r}AB} \right)=  8\pi \left[ \left( 
P+\rho \right)a^2A^4 u^2  +P \right] 
\,,    \label{delta1} \\  
&& \nonumber \\
&& -\left( \frac{A}{B} \right)^2 \left( 2\dot{C}+3C^2 
+ \frac{2\dot{m}C}{\bar{r}AB} \right)=8\pi  P  \,.
\label{delta2} 
\end{eqnarray}
Eqs.~(\ref{delta1}) and (\ref{delta2}) combined  
give $P=-\rho$: only the de Sitter equation of 
state is allowed and then eq. (\ref{delta0}) implies that  
$\dot{M}=0$.   

\subsubsection{Imperfect fluid and no radial mass flow}
  
Consider now  the imperfect fluid stress-energy tensor 
\begin{equation} \label{imperfect}
T_{ab}=\left( P+\rho \right)u_a u_b +P g_{ab}+ q_a u_b+q_b u_a 
\,,
\end{equation}
as a source for the generalized McVittie solutions, where 
the purely spatial vector $q^c$ describes a radial energy flow, 
\begin{equation} 
u^{\mu}=\left( \frac{A}{B}, 0,0,0 \right) \,, \;\;\;\;\;
q^{\alpha}=\left( 0, q, 0,0 \right) \;, \;\;\;\;\; q^cu_c=0 
\,,
\end{equation}
and $ u^c u_c=-1 $ (in principle one could take $q^c$ to be 
spacelike instead of purely spatial \cite{CarreraGiuliniPRD10, 
Anderson11}). The $\left( t, \bar{r}\right)$ component of the 
Einstein equations yields
\begin{equation} 
\frac{\dot{m}}{m}+\frac{\dot{a}}{a}= -\frac{4\pi G}{m}\, 
\bar{r}^2 a^2 A^4 B^2 q  \,.
\end{equation}
Furthermore, it is 
\begin{equation} 
\frac{\dot{M}}{M}=\frac{\dot{m}}{m}+\frac{\dot{a}}{a}
\end{equation}
and the area of a sphere $\Sigma$ of constant time and 
constant isotropic  radius $\bar{r}$ is 
\be
{\cal A}=\int \int d\theta d\varphi \, 
\sqrt{g_{\Sigma}}=4\pi a^2 A^4 \bar{r}^2 \,,
\ee
then energy flow, area ${\cal A}$, and accretion rate are related by  
\begin{equation}  \label{accretionrate}
\dot{M} (t)=- a B^2 {\cal A} q  \,.
\end{equation}
In the case of  inflow ($q<0$), this condition 
can be written on a sphere of radius $\bar{r} \gg m$ as 
$\dot{M} \simeq  a {\cal A} \left| q  \right| $; 
for a 2-sphere, $M$ increases due to the  inflow of 
matter alone (but it receives another contribution from the 
evolution of the cosmological fluid contained in it).

The energy density and pressure obtained from the  Einstein equations are
\begin{eqnarray}
\rho \left( t, \bar{r}\right) &=&
\frac{1}{8\pi } \, \frac{3A^2}{B^2} \left( 
\frac{\dot{a}}{a}+\frac{\dot{m}}{\bar{r}A} \right)^2 \,, 
\label{newdensity} \\
&&\nonumber \\
P \left( t, \bar{r}\right) &=&
\frac{- A^2}{8\pi  B^2} \left\{ 
2 \frac{d}{dt} \left( \frac{\dot{a}}{a}+\frac{\dot{m}}{\bar{r}A} 
\right) + 
\left( \frac{\dot{a}}{a}+\frac{\dot{m}}{\bar{r}A} 
\right)\left[ 3
\left( \frac{\dot{a}}{a}+\frac{\dot{m}}{\bar{r}A} 
\right)+\frac{2\dot{m}}{\bar{r}AB} \right]\right\}\,; 
\label{newpressure}
\end{eqnarray}
clearly  the energy density is always non-negative. 
In terms of the quantity $C$, eq. (\ref{newpressure}) 
becomes the generalization of the  Raychaudhuri equation of FLRW space 
\begin{equation} \label{generalizedRaychaudhuri}
\dot{C}=-\,\frac{3C^2}{2}-\frac{\dot{m}}{\bar{r}AB}\, C -4\pi  
\,  \frac{B^2}{A^2} \, P \,.
\end{equation}
It reduces to the usual Raychaudhuri equation of FLRW cosmology in the 
limit $m\rightarrow 0$,  
\be
\dot{H}=-\frac{3H^2}{2}-4\pi  P \,,
\ee
and then the  Hamiltonian 
constraint $H^2=8\pi  \rho/3$ yields 
\begin{equation} \label{FLRWHdot}
\dot{H}=-4\pi  \left(P+\rho \right) \,.
\end{equation}
When $m\neq 0$, instead, eq. (\ref{newdensity}) 
yields the generalization \cite{AudreyPRD} 
\begin{equation} 
\dot{C}=-4\pi  \, \frac{B^2}{A^2}\left( P+\rho \right) 
- \frac{\dot{m}C}{\bar{r}AB} \,.
\end{equation}

\subsubsection{Imperfect fluid and radial mass flow}

Let us consider now an imperfect fluid with 
stress-energy tensor of  the form (\ref{imperfect}) with  
both radial mass flow and energy current present 
and of the form 
\begin{equation} 
 u^{\mu}=\left( \frac{A}{B}\sqrt{1+a^2A^4u^2}, u, 0, 0  
\right)\,, \;\;\;\;\;\;\;\;\;
q^{\mu}=\left( 0,q,0,0 \right) \,.
\end{equation}
By using the components (\ref{einst1})-(\ref{einst3}) of the 
Einstein tensor, the field equations become
\begin{equation}  \label{deltadelta1}
\dot{M} =- aB^2 {\cal A} \sqrt{1+a^2A^4u^2} \left[ \left( 
P+\rho 
\right)u+q \right] \,,
\end{equation}
\begin{equation} 
-3\left( \frac{AC}{B} \right)^2 =-8\pi \left[ \left( P+\rho 
\right)a^2 A^4 u^2 +\rho \right] \,,
\end{equation}
\begin{equation} 
-\left( \frac{A}{B} \right)^2 \left( 2\dot{C}+3C^2 
+\frac{2\dot{m}C}{\bar{r}AB} \right)=8\pi \left[ \left( P+\rho 
\right)a^2 A^4 u^2 +P+2a^2 A^4 qu \right] \,,
\end{equation}
\begin{equation} 
-\left( \frac{A}{B} \right)^2 \left( 2\dot{C}+3C^2 
+\frac{2\dot{m}C}{\bar{r}AB} \right)=8\pi P \,.
\end{equation}
Adding the last two equations yields  
\begin{equation} \label{460}
q=-\left(P+\rho \right)\frac{u}{2} 
\end{equation}
(equivalently, this equation can be seen as a consequence of the spatial 
Ricci isotropy $G^{\bar{r}}_{\bar{r}} =G^{\theta}_{\theta}$), 
{\em i.e.}, to an ingoing radial mass flow 
there corresponds an outgoing radial heat current if $P>-\rho$. By 
substituting eq. (\ref{460}) into eq. (\ref{deltadelta1}), one obtains 
the accretion rate
\begin{equation} 
\dot{M} =-\frac{1}{2} aB^2 \sqrt{1+a^2A^4 u^2} \left( P+\rho 
\right){\cal A}u \,,
\end{equation}
where $\left( P+\rho \right) {\cal A} u $ can be seen as the 
flux of gravitating energy through the surface of area ${\cal 
A}$ (remember that  $u<0$). The energy density is given by
\begin{equation} 
8\pi  \rho= \frac{A^2}{B^2}\left[ 3C^2 +\left( 
\dot{C}+\frac{\dot{m}C}{\bar{r}AB} \right) 
\frac{2 a^2A^4u^2}{1+a^2A^4 u^2} \right] \,.
\end{equation}

\subsubsection{The ``comoving mass'' solution}

In the class of generalized McVittie solutions of GR, the 
choice $M(t)=m_0 \, a(t)$ where $m_0$ is a constant, 
selects a special one which is 
a late-time attractor within this class. The corresponding line 
element in isotropic coordinates is 
\be
\label{eq:ds-ph}
ds^2 = -\frac{\left(1-\frac{m_0}{2r}\right)^2}{
\left(1+\frac{m_0}{2r}\right)^2} \, dt^2
  +{a^2\left(t\right)}\left(1+\frac{m_0}{2r} \right)^4
\left(dr^2+r^2d\Omega^2_{(2)} \right) \,.
\ee  
The apparent horizons of this metric were studied in Ref. 
\cite{GaoChenFaraoniShen08} by transforming to areal radius. A nice feature 
of this solution is that the apparent horizons are given analytically by 
\begin{eqnarray}
R_{c}&=&{\frac{1}{2H}}{\left(1+\sqrt{1-8m_0\dot{a}} \, 
\right)} \,, \\ && \nonumber\\
R_{b}&=&{\frac{1}{2H}}{\left(1-\sqrt{1-8m_0\dot{a}} \, 
\right)} \,. \label{Rb}
\end{eqnarray}
$R_{c}$ is a cosmological and
$R_{b}$ is a black hole apparent horizon.  
The surface $r=m_0/2$ (or $ \tilde{r}=2m_0 $, or 
$ R=2m_0a=2M(t) $) is a spacetime singularity 
contained inside the black  hole apparent horizon when the 
latter exists, since   $R_{c,b} > 2m_0 a 
=2M$ \cite{GaoChenFaraoniShen08}. The black hole and cosmological apparent 
horizons have the qualitative behaviour already discussed 
for all the McVittie and generalized McVittie solutions 
\cite{GaoChenFaraoniShen08}.

\subsubsection{The general class of solutions}

For  the wider class of 
generalized  McVittie solutions with arbitrary dependence 
$m(t) \geq0$, an analysis of the apparent horizons using the areal radius 
\cite{GaoChenFaraoniShen08} identifies them as the roots of the equation 
\begin{equation} 
\label{addhorizons} 
HR+\dot{m}a\sqrt{
\frac{\tilde{r}}{r} } =\pm \left( 1-\frac{2M}{R} \right) 
\,,
\end{equation}
where  $\tilde{r}\equiv R/a$.  
Since $M(t)=m(t)a(t)$, the left hand side can be written as
 \begin{equation} 
HR+ M\left( 1+\frac{m}{2r} \right)\left(
\frac{\dot{M}}{M}-H \right) \, 
\end{equation} 
where the factor $M\left( 1+\frac{m}{2r} \right)$ quantifies  
the deviation
of the radius from $2M$ ($ r>m/2$ corresponds to $  
R>2M$ and to $M\left( 1+\frac{2m}{r} \right) >2M$), while 
the factor $ \left( \frac{\dot{M}}{M}-H \right) $ is the 
difference between the percent rate of change of $M$ and 
that of the 
scale factor of the substratum. The  vanishing of this
factor corresponds to an analog of 
stationary accretion for a time-dependent background. Then,
the special solution with $M(t)=m_0 a(t)$  corresponds to 
stationary accretion relative to the FLRW background.

Eq. (\ref{addhorizons}), which  
becomes
\begin{equation}\label{addquadratic} 
HR^2 +\left[ M\left( 1+\frac{m}{2r} \right)\left( 
\frac{\dot{M}}{M}-H \right)-1 \right]R +2M=0 \,, 
\end{equation} 
is not a quadratic algebraic equation, but it can be treated formally
as such, providing the formal roots 
\be
 R_{c,b}=\frac{1}{2H} \left\{ 1-M\left( 1+\frac{m}{2r}
\right)\frac{\dot{m}}{m} \pm \sqrt{ \left[ 1-M\left( 
1+\frac{m}{2r}
\right)\frac{\dot{m}}{m} \right]^2-8m\dot{a} } \right\} \,.
\ee 
Since $r=r(R)$, this is really an implicit 
equation for the radii $R_{c,b}$ of the cosmological and 
black hole apparent horizons. When the argument
of the square root is positive there are a cosmological 
apparent horizon  at $R_c$ and a black hole apparent 
horizon at $R_b$. When this argument vanishes, these two apparent 
horizons coincide at $ \sqrt{\frac{2M}{H}}$. 
If this argument becomes negative, the apparent horizons 
disappear leaving behind a naked singularity \cite{GaoChenFaraoniShen08}.

\subsubsection{Attractor behaviour of the ``comoving mass'' 
solution}

``Comoving mass'' solutions are generic under certain assumptions,
 in the sense that all 
other generalized McVittie solutions  approach 
them at late times \cite{FaraoniGaoChenShen09}. In fact, assume
that the universe always expands, that $m(t) \geq 0$, and 
that the function $m(t)$ is continuous with its first 
derivative. Then, using  $\tilde{r}\equiv R/a$, one obtains    
\be\label{200}
H \tilde{r}+ \frac{2m}{\tilde{r} a} = -\dot{m}\left( 
1+\frac{m}{2r}  \right) + \frac{1}{a} 
\,.
\ee
Since $m\geq 0$ the left-hand side is  always non-negative 
and  $\dot{m} \left( 1+\frac{m}{2r}\right)<\frac{1}{a}$. 
Then, given that  $1+\frac{m}{2r}>0$, in  an expanding  
universe in which $ a\rightarrow + \infty$,  one has  $ 
\dot{m}_{\infty} \equiv \lim_{t\rightarrow 
+\infty} \dot{m}(t) \leq 0$. If  $\dot{m}_{\infty}=0$, the 
quantity $m(t)$ 
becomes asymptotically comoving.

The other  possibility is  $\dot{m}_{\infty}<0$. In this 
case, there is a time $\bar{t}$ such that 
$\forall \, t>\bar{t}$ it is $\dot{m}(t)<0$. Then there are 
only two options: since $ m(t) \geq 0$, 
either $m(t)$ reaches the value zero at a finite time $t_*$ 
with  derivative  $\dot{m}_*\equiv \dot{m}(t_*) < 0$, or
else  $m(t) \rightarrow m_0=$const. with 
$\dot{m}(t)\rightarrow 0$, {\em 
i.e.}, $m(t)$ has a horizontal asymptote.

In the first case one has, at $t=t_*$, ~$ HR =\left| 
\dot{m}_*\right| a +1 $,  which yields the radius of the 
black hole apparent horizon at $t_*$
\be
r_* \equiv r_{horizon}(t_*)= \frac{1}{H(t_*)} 
\left(|\dot{m}_{*}| +\frac{1}{a} \right) \;.
\ee
Late in the history of the universe we have  a black hole 
of zero mass $ M(t_*)=a(t_*)m(t_*)$ but finite 
radius $r_*$. As time 
evolution  continues, one would have negative mass $M$ and 
finite radius of the black hole apparent horizon. This unphysical
situation  for $m(t_*)=0$ with $m(t>t_*)<0$ is discarded.

In the second  case  $\dot{m}(t) 
\rightarrow 0$ at late 
times and  $t\rightarrow +\infty$ if the cosmic 
expansion  continues forever or $ t\rightarrow t_{rip} $ 
if a Big Rip occurs at $t_{rip}$. The physical 
meaning of $\dot{m} \rightarrow 0$ is that, at late times, 
the rate of increase of the black hole mass  is at most the 
Hubble rate and becomes comoving.

\subsection{The Sultana-Dyer solution}

\begin{figure}
\centering
\includegraphics[width=8.5cm]{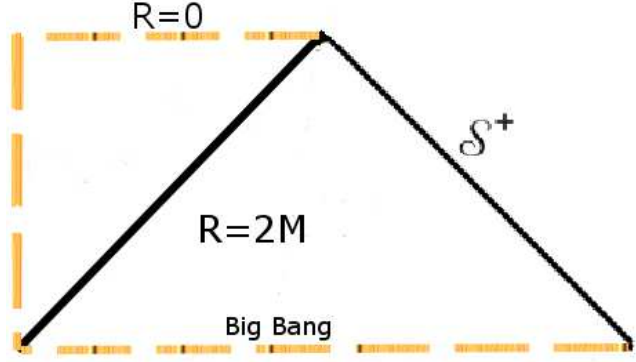}
\caption{Conformal diagram of the Sultana-Dyer spacetime. 
\label{SultanaDyerReview}}
\end{figure}

The Sultana-Dyer solution of GR \cite{SultanaDyer05} is a Petrov type~D 
metric interpreted as a black hole embedded in a spatially 
flat FLRW universe. This solution was 
generated by extending a metric resulting from 
the conformal  transformation of 
the Schwarzschild metric  $g_{ab}^{(S)} \rightarrow 
\Omega^2 \, g_{ab}^{(S)}$ with conformal factor  
$\Omega=a(t)=\eta^2$ equal to the scale factor of a 
dust-filled $k=0$ FLRW universe in conformal time 
$\eta$. That is, this spacetime is 
conformally static and admits a  conformal Killing vector 
$\xi^a$ (fig.~\ref{SultanaDyerReview}).

The authors of \cite{SultanaDyer05} aimed at changing the Schwarzschild  
timelike Killing field  $\xi^c$ into a  conformal Killing 
field defined for $\xi^c  \nabla_c \Omega \neq 0$, 
thus generating a conformal Killing horizon (which, however, 
seems of little relevance  in modern studies of time-evolving horizons).

The Sultana-Dyer metric is 
\begin{equation} \label{SultanaDyeroriginal}
ds^2=a^2(\eta) \left[ 
- \left(1-\frac{2m_0}{r} \right)d\eta^2 +\frac{4m_0}{r} \, 
d\eta dr 
+  \left(1+\frac{2m_0}{r} \right)dr^2 +
r^2 d\Omega^2_{(2)} \right] \,,
\end{equation} 
where $m_0$ is a constant and $a(\eta)=\eta^2$. 
The coordinate transformation
\be
\eta (t, r) =t+2m_0 \ln \left| \frac{r}{2m_0} -1 \right| 
\,,
\ee
turns the line element into the form
\begin{equation} 
ds^2=a^2(t, r) \left[ 
- \left(1-\frac{2m_0}{r}\right) dt^2 
+ \frac{dr^2}{ 1 - \frac{2m_0}{r} } +
r^2 d\Omega^2_{(2)} \right] \,,
\end{equation} 
which is explicitly conformal to the Schwarzschild metric 
with  conformal factor
\be
\Omega = a(t, r)=\eta^2 (t,r)=
\left( t+2m_0 \ln \left| \frac{r}{2m_0} -1 \right| 
\right)^2 \,.
\ee
The matter source of the Sultana-Dyer spacetime is a  
mixture of two non-interacting perfect fluids with stress-energy tensor 
\be
T_{ab}= T_{ab}^{(I)}+T_{ab}^{(II)} \,,
\ee
where $ T_{ab}^{(I)}=\rho u_a\, u_b $ describes an ordinary 
dust with timelike 4-velocity $u^c$ and $ 
T_{ab}^{(II)}=\rho_n \,  k_a\, k_b$  describes a null dust 
with density $\rho_n $ and $k^c k_c=0$ 
\cite{SultanaDyer05}.  A problem of this  
solution is that the cosmological fluid 
becomes tachyonic with negative  energy density at late 
times near $\bar{r}=m_0/2$ 
\cite{SultanaDyer05}.

Let us use, in the rest of this subsection, the quantity 
\be\label{bigM}
 M(\bar{t}) \equiv m_0 \, a( \bar{t}) \,,
\ee
which is not constant in the Sultana-Dyer solution. The
locus $r=2m_0 $ is not a singularity, but the conformal 
factor $\Omega$ vanishes there. The metric 
(\ref{SultanaDyeroriginal}), however, is not singular 
there. The Ricci curvature is 
\be
{R^a}_a= \frac{12}{\eta^6} \left( 1-\frac{2m_0}{r} 
+\frac{2m_0\eta}{r^2} \right) \,,
\label{ssscurvature} 
\ee 
and is not singular at $r=2m_0 $ (where $\eta\rightarrow 
-\infty$) but is singular at $r=0$ (central singularity)  
and for $\eta=0$ (Big Bang singularity).

The problem of Hawking emission from the Sultana-Dyer black hole was 
approached in Ref. \cite{SaidaHaradaMaeda07}. These authors 
considered quantum radiation from a massless conformally 
coupled scalar field $\phi$ and computed the renormalized 
stress-energy tensor $\langle T_{ab} \rangle $ of $\phi$ 
taking advantage of the simplifications introduced by the 
fact that the Sultana-Dyer spacetime is conformal to the 
Schwarzschild one and taking into account the conformal 
anomaly and  particle creation by the FLRW background. 
Discarding complicated corrections which are small if the black hole is 
evolving slowly, the effective Hawking temperature  from the 
Sultana-Dyer black hole was computed as 
\cite{SaidaHaradaMaeda07}
\be
T_{eff}= \frac{1}{8\pi m_0 a(t)} = \frac{T_{Schw}}{a(t)} 
\,,
\ee
where $T_{Schw}=(8\pi m_0)^{-1}$ is the Hawking temperature 
of the Schwarzschild black hole which seeds the 
Sultana-Dyer metric. The more general relation 
\be\label{Tconformal}
T=\frac{ T_{Schw}}{\Omega}
\ee
for spacetimes conformally related to the Schwarzschild 
spacetime by a trasformation with conformal factor $\Omega$ 
is conjectured in \cite{SaidaHaradaMaeda07}. Independent 
support for eq. (\ref{Tconformal}) comes from dimensional 
considerations related to the use of conformal transformations 
\cite{myHawkingT}.

\subsection{The Husain-Martinez-Nu\~nez solution}

In the 1994 Husain-Martinez-Nu\~nez solution of GR 
 \cite{HusainMartinezNunez} a  new 
phenomenology of the apparent 
horizons appears. This spacetime  describes an inhomogeneous universe with 
a spatially flat FLRW background sourced by a free, 
minimally coupled, scalar field. The coupled Einstein-Klein-Gordon equations 
reduce to
\begin{eqnarray}
&& R_{ab}=8\pi \nabla_a \phi  \nabla_b \phi \,.
\,, \label{HMNfieldeq1}\\
&&\nonumber\\
&& \Box \phi=0 \,,
\end{eqnarray}
and the Husain-Martinez-Nu\~nez solution to them is 
\cite{HusainMartinezNunez}  
\begin{eqnarray}
ds^2 &=& \left( A_0 \eta +B_0 \right) \left[ - \left( 
1-\frac{2C}{r}\right)^{\alpha} 
d\eta^2 +\frac{dr^2}{\left( 1-\frac{2C}{r}\right)^{\alpha} 
} \right. \nonumber\\
&&\nonumber\\
&\, & \left. + r^2 \left( 
1-\frac{2C}{r}\right)^{1-\alpha} 
d\Omega_{(2)}^2 \right] \,,  \label{HMNconformaltime}\\
&&\nonumber\\
\phi(\eta, r ) &=& \pm \frac{1}{4\sqrt{\pi}} \, \ln \left[ 
D\left( 1-\frac{2C}{r}\right)^{\alpha/\sqrt{3}} 
\left( A_0 \eta +B_0 \right)^{\sqrt{3}} \right] 
\,,\label{HMNscalar}
\end{eqnarray}
where $A_0 , B_0 , C$, and $D$ are non-negative constants, 
$\alpha =\pm \sqrt{3}/2$, and $\eta >0$. 
The  additive constant $B_0$ becomes irrelevant and can be 
dropped whenever $A_0\neq 0$. When $A_0=0$, the 
Husain-Martinez-Nu\~nez metric degenerates into the 
static Fisher  spacetime \cite{Fisher48}
\be \label{Fisher}
ds^2=-V^{\nu}(r) \, d\eta^2 +\frac{dr^2}{V^{\nu}(r)} +r^2 
V^{1-\nu}(r) d\Omega_{(2)}^2 \,,
\ee
where $V(r)=1-2\mu/r$, $\mu$ and $\nu$ are parameters,  and 
the Fisher scalar field is 
\be 
\psi(r)=\psi_0 \ln V(r) \,.
\ee
The Fisher solution of the coupled Einstein-Klein-Gordon equations, 
also referred to as the Janis-Newman-Winicour-Wyman 
solution, has been rediscovered many times 
\cite{BergmanLeipnik57, JanisNewmanWinicour68, 
Buchdahl72, Wyman81, AgneseLaCamera85, Virbhadra97}. Its features are  
a naked singularity at $r=2C$ 
and its  asymptotic flatness. It is claimed that this solution is the most 
general static and spherically symmetric solution of the 
Einstein equations with zero cosmological constant and a 
massless, minimally coupled, scalar field 
\cite{Roberts93}, but it is unstable \cite{Abe88}. The 
general Husain-Martinez-Nu\~nez metric is 
conformal to the Fisher metric with conformal factor 
$\Omega=\sqrt{A_0 \eta+B_0}$ equal to the scale factor 
of the background FLRW space and with only two possible 
values of the parameter $\nu$. From now on, we set the 
constant  $B_0$ to zero by labelling the Big Bang by  
 $\eta=0$.  
The sign in eq. (\ref{HMNscalar}) is not 
associated with the sign of $\alpha$. The full metric 
is asymptotically FLRW for $r\rightarrow +\infty$ and is 
FLRW if $C =0$ (in which case the constant $A_0$ can be 
eliminated by rescaling the time coordinate $\eta$).

\begin{figure}
\centering
\includegraphics[width=8.5cm]{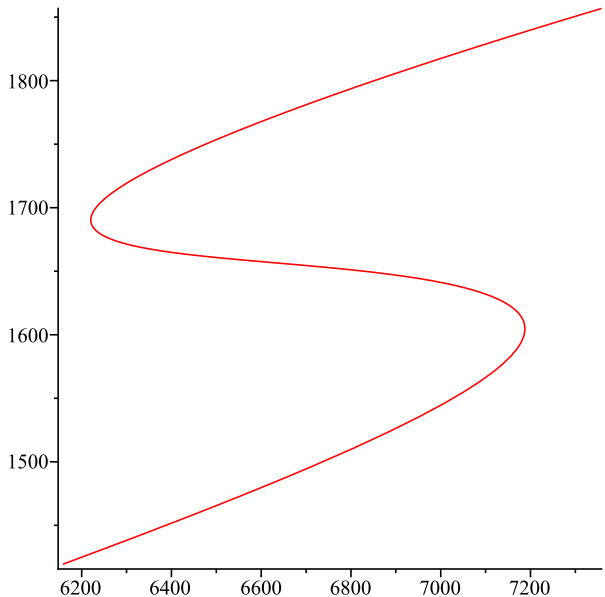}
\caption{The radii of the apparent horizons of the 
Husain-Martinez-Nu\~nez spacetime (vertical axis)  
versus comoving time (horizontal axis) 
for  $\alpha=\sqrt{3}/2$ ($t$ and $R$ are 
measured in arbitrary units of length  and the 
parameter values  are chosen so that 
$(Ca_0)^{3/2}=10^3$ in eq. (\ref{HMNparametricR})). 
\label{HMNfigure1}}
\end{figure}

\begin{figure}
\centering
\includegraphics[width=8.5cm]{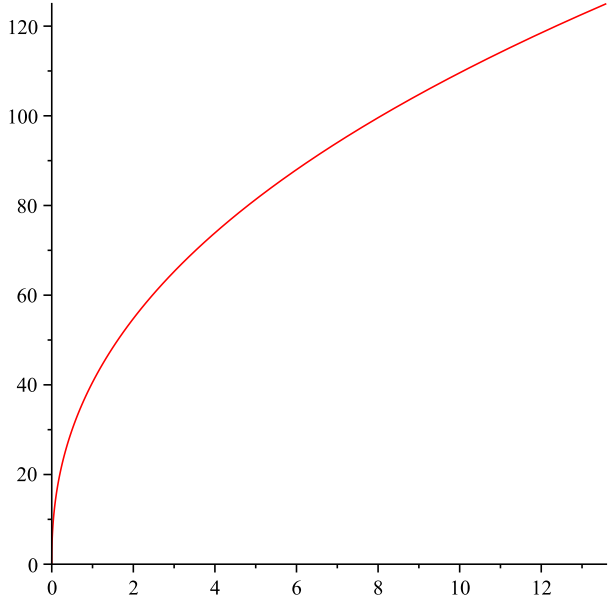}
\caption{The radius of the Husain-Martinez-Nu\~nez 
apparent horizon (vertical axis) versus comoving time (horizontal axis) 
for 
$\alpha=-\sqrt{3}/2$. There is always only one, expanding, cosmological 
apparent horizon and there is a naked singularity at $R=0$.
 \label{HMNfigure2}}
\end{figure}

The Ricci scalar 
\be
{R^a}_a = 8\pi \nabla^c\phi \nabla_c \phi = 
\frac{2\alpha^2 C^2 \left( 
1-\frac{2C}{r}\right)^{\alpha-2}}{
3 r^4  A_0 \eta } 
- \frac{ 3A_0^2}{ 
2 \left( A_0 \eta \right)^3  
\left( 1-\frac{2C}{r}\right)^{\alpha} } \,,
\ee
immediately identifies a spacetime singularity at 
$r=2C$ (for both values of the 
parameter $\alpha$). The scalar  $\phi$ also diverges there, and a 
Big Bang singularity is present at $\eta =0$. Only the 
coordinate range $ 2C <r<+\infty $ is physical and the lower limit
 $r=2C$ corresponds to zero areal radius
\be\label{HMNarealradius}
R(\eta, r)= \sqrt{A_0 \eta} \, r \left( 
1-\frac{2C}{r}\right)^{\frac{1-\alpha}{2}} \,.
\ee
Let us introduce the comoving time $t$ defined 
by  $ dt=ad\eta$ (where $a(\eta)=\sqrt{A_0\eta}$ is the 
FLRW scale factor) in place of  the conformal time $\eta $, then it is  
\be
t=\int d\eta \, a(\eta)=\frac{2\sqrt{A_0}}{3} \, \eta^{3/2} 
\ee 
by choosing $\eta=0$ at $t=0$, or 
\be
\eta=\left( \frac{3}{2\sqrt{A_0}} \, t \right)^{2/3}
\ee
and
\be
a(t)=\sqrt{A_0\eta}= a_0 \,  t^{1/3} \,, \;\;\;\;\; 
a_0=\left( \frac{3A_0}{2} \right)^{1/3} \,.
\ee
This power law for the scale factor is consistent with the 
stiff equation of state 
$P=\rho/3$ of a free massless scalar field in a  FLRW 
universe and with the general solution 
$ a(t)= \mbox{const.} \,  t^{\frac{2}{3(w+1)}}  $ (where $w\equiv P/\rho$).
The Husain-Martinez-Nu\~nez solution in comoving time reads
\be
ds^2 = - \left( 1-\frac{2C}{r}\right)^{\alpha} 
dt^2 +a^2(t) \left[ \frac{ dr^2}{\left( 
1-\frac{2C}{r}\right)^{\alpha} } 
+ r^2 \left( 1-\frac{2C}{r}\right)^{1-\alpha} 
d\Omega_{(2)}^2 \right]  \label{HMNcomoving}
\ee
with 
\be
\phi( t, r ) = \pm \frac{1}{4\sqrt{\pi}} \, \ln \left[ 
D\left( 1-\frac{2C}{r}\right)^{\alpha/\sqrt{3}} 
a^{2\sqrt{3}}(t) \right] \,.
\ee
The areal radius (\ref{HMNarealradius})  
increases with $r$ for 
$ r>2C$. It is useful to rewrite the line element  
in terms of the areal radius $R$. By setting
\be
A(r) \equiv 1-\frac{2C}{r} \,, \;\;\;\;\;\; 
B(r) \equiv 1-\frac{(\alpha+1) C}{r} \,, 
\ee
we have $R(r)=a(t)r A^{\frac{1-\alpha}{2}}(r) $
and 
\be
dr=\left[ A^{\frac{\alpha +1}{2}} \frac{dR}{a} -AH \, rdt 
\right] \frac{1}{B(r)} \,.
\ee
The metric is then  
\begin{eqnarray} 
ds^2&=& -A^{\alpha} \left[ 1- \frac{ 
H^2R^2 A^{2(1-\alpha)} }{B^2(r)} \right] dt^2 
+\frac{H^2R^2A^{2-\alpha}(r) }{B^2(r)}\, dR^2 \nonumber\\
&&\nonumber\\
&\, & - \, \frac{2HR A^{\frac{3-\alpha}{2}}}{B^2(r)} \, 
dt \, dR 
 +R^2 d\Omega_{(2)}^2 \,.\label{HSMcross}
\end{eqnarray}
The time-radius cross-term is eliminated by introducing  
a new time  $T$ with differential  
\be 
dT= \frac{1}{F} \left( dt+\beta dR \right) \,,
\ee
where $\beta(t, R)$ is a function to be determined and 
$F(t, R)$ is an integrating factor which must satisfy 
\be
\frac{\partial}{\partial R} \left( \frac{1}{F} \right)= 
\frac{\partial}{\partial t} \left( \frac{\beta}{F} \right)
\ee
in order for $dT$ to be an exact differential. Using 
$dt=FdT-\beta dR$ in eq.~(\ref{HSMcross}) and choosing 
\be 
\beta(t,R)= \frac{ HRA^{\frac{3(1-\alpha)}{2}} }{
B^2(r)-H^2R^2 A^{2(1-\alpha)} } 
\,,
\ee 
the  line element becomes
\begin{eqnarray}
ds^2 &=&- A^{\alpha}(r) \left[ 
1- \frac{H^2R^2 A^{2(1-\alpha)}(r)  }{B^2(r)} \right] 
F^2dt^2 \nonumber\\
&&\nonumber\\
&\, & +\frac{H^2 R^2 A^{2-\alpha}(r)}{B^2(r)} \left[ 1+  
\frac{  A^{1-\alpha}(r)} { B^2(r) -H^2R^2  
A^{2(1-\alpha)}(r)}
\right] dR^2 +R^2 d\Omega_{(2)}^2 \,.\nonumber\\
&&
\end{eqnarray}
The apparent horizons, located by  
$g^{RR}=0$, must satisfy
\be\label{HMNAH}
B(r)=H(t)RA^{1-\alpha}(r) \,,
\ee
where now $r=r(t,R)$, or
\be
\frac{1}{\eta} = \frac{2}{r^2} \Big[ r-(\alpha+1)C \Big] 
\left( 1-\frac{2C}{r} \right)^{\alpha -1}  
\label{HMNoriginalAH}
\ee
using the original coordinates 
$\left( \eta, r\right)$ \cite{HusainMartinezNunez}. 
For $r\rightarrow +\infty$ (corresponding to $R\rightarrow 
+\infty$), this equation reduces to $R\simeq H^{-1}$, the 
radius of the cosmological apparent horizon in 
spatially flat FLRW space. Eq. (\ref{HMNAH})  must be 
solved numerically. 
Let $x\equiv C/r$, then the equation locating the apparent 
horizons is
\be \label{TeKaPo}
HR=\left[ 1-\frac{(\alpha+1)C}{r}  \right] \left( 
1-\frac{2C}{r} \right)^{\alpha-1} \,.
\ee
The left hand side can be written as 
\be
HR=\frac{a_0}{3 \, t^{2/3}} \, \frac{2C}{x} \left( 1-2x 
\right)^{\frac{1-\alpha}{2}} \,,
\ee
which expresses the radius of the apparent 
horizons in units of $H^{-1}$ (the radius of 
the cosmological apparent horizon of the FLRW 
background if it  did not have the central 
inhomogeneity). The right 
hand side is $\left[ 
1-(\alpha+1)x\right](1-2x)^{\alpha-1} $. Eq. (\ref{TeKaPo}) 
and the equation defining the areal radius give
\begin{eqnarray}
t(x) &=& \left\{ \frac{2Ca_0}{3} \, \frac{ 
(1-2x)^{3(1-\alpha)}}{x\left[ 1-(\alpha+1)x\right]} 
\right\}^{3/2} \,,\label{HMNparametricR}\\
&&\nonumber\\
R(x) &=& a_0 \, t^{1/3}(x) \, \frac{2C}{x}\left( 1-2x 
\right)^{ \frac{1-\alpha}{2}} \,.
\end{eqnarray}
This is a parametric 
representation of the function $R(t)$ and can be used to 
plot this function. The result is illustrated in 
figs. \ref{HMNfigure1} and \ref{HMNfigure2}.
If  $\alpha=\sqrt{3}/2$, between the Big Bang and 
a critical time $t_*$ there is 
only one expanding apparent horizon, then two other 
apparent horizons are created at $t_*$. One is a 
cosmological  apparent horizon which expands forever and the other is a 
black hole horizon which contracts until it meets the first 
(expanding) black hole apparent horizon \cite{HusainMartinezNunez}. 
When they meet, these 
two annihilate and a naked singularity appears at 
$R=0$ in a FLRW universe. 
This phenomenology of apparent horizons differs from 
that of the McVittie and generalized McVittie solutions. The  
``S-curve'' phenomenology of fig.~\ref{HMNfigure1} 
appears also in Lema\^itre-Tolman-Bondi  spacetimes already 
for a  dust fluid much simpler than a scalar field  
\cite{BoothBritsGonzalezVDB} 
(multiple ``S''s are possible, for example  
see fig.~9 
of Ref.~\cite{BoothBritsGonzalezVDB}) 
and in analytical solutions of  
Brans-Dicke and $f\left({R^c}_c \right)$ gravity. 
The scalar field is regular on the apparent horizons.

For $\alpha=-\sqrt{3}/2$ there is only one, forever expanding,  cosmological 
apparent horizon and the universe 
contains a naked singularity at $R=0$ (fig. 
\ref{HMNfigure2}), with the usual Big Bang singularity 
at $t=0$.

The apparent horizons are spacelike \cite{HusainMartinezNunez}, 
as can be seen by studying  
the normal vector to these surfaces and checking that 
it always lies inside the light cone in an $\left( 
\eta, r \right)$ diagram.
Eq. (\ref{HMNoriginalAH}) yields 
\be
\eta = \frac{r^2 \left(1-\frac{2C}{r} 
\right)^{1-\alpha}}{2\left[ r-C(1+\alpha) \right]}
\ee
along the apparent horizons. Differentiate this 
relation with respect to $R$ to obtain
\be
\eta_{, r} \Big|_{AH} = \left( 1-\frac{2C}{r} 
\right)^{-\alpha} 
\left\{ 1-\frac{r^2 \left(1-\frac{2C}{r} \right)}{
2\left[ r-C(1+\alpha) \right]^2} \right\}\,.
\ee
Along radial null geodesics it is 
\be
\eta_{, r}\Big|_{light\, cone} = \pm \left( 1-\frac{2C}{r} 
\right)^{-\alpha} \,,
\ee
which follows from $ds^2=0$ with 
$d\theta=d\varphi=0$. Therefore, it is 
\cite{HusainMartinezNunez}
\be
\left| \frac{ 
\eta_{, r}\Big|_{AH} }{
\eta_{, r}\Big|_{light\, cone} } \right| =1- 
\frac{\left( 1-\frac{2C}{r} \right)}{
2\left[ 1-\frac{(\alpha +1)C}{r} \right]^2} \leq 1 
\ee
and the normal to the apparent horizons is always 
pointing inside the light cone, except at the spacetime points at which this 
vector becomes tangent to the light cone and is null, which 
occurs when a pair of apparent horizons is created or destroyed 
\cite{HusainMartinezNunez}. This occurrence is in agreement with a general 
result of Ref.~\cite{BoothBritsGonzalezVDB} stating that a trapping horizon created 
by a massless scalar field 
must be spacelike (however, even simple potentials $V(\phi)$ can make the trapping 
horizon be non-spacelike).

The nature of the singularity at $r=2C$ (or $R=0$) is easily established: 
all surfaces  $R=$const. have 
equation 
$\Phi (R) \equiv R-\mbox{const.}=0$ and  gradient
$N_{\mu} \equiv \nabla_{\mu} \Phi=\delta_{\mu R}$ in 
coordinates $\left( t, R, \theta, \varphi \right)$. The norm 
squared is 
\be
N_c N^c= g^{RR}=\frac{B^2}{H^2R^2A^{2-\alpha}} \, 
\frac{1}{
1+\frac{A^{1-\alpha}}{B^2-H^2R^2A^{2(1-\alpha)}}} 
\ee
and, because $B(r) \rightarrow \frac{1-\alpha}{2}$ and 
$A(r)\rightarrow 0^{+}$ as $r\rightarrow 2C^{+}$, it is 
$N_cN^c>0$ and $N_cN^c \rightarrow +\infty$ as 
$r\rightarrow 2C^+$. The singularity at $R=0$ is 
timelike for both values of the parameter $\alpha$.

For $\alpha=+\sqrt{3}/2$, the Husain-Martinez-Nu\~nez spacetime is interpreted 
as describing the creation 
and annihilation of pairs of black hole apparent horizons. 
The central singularity at $R=0$ is created with the 
universe in the Big Bang and does not result from a  
collapse process (this is also the case for 
$\alpha=-\sqrt{3}/2$).

\subsection{The Fonarev and generalized Fonarev solutions}

The Fonarev solution of the Einstein 
equations with a minimally coupled scalar field in an  
exponential potential as the matter source \cite{fo:1995} generalizes the 
Husain-Martinez-Nu\~nez solution. 
It describes a central inhomogeneity embedded 
in a scalar field FLRW universe.  The action is 
\begin{eqnarray}
 S=\frac{1}{2\kappa} \int d^4x\sqrt{-g}\left[ {R^a}_a -\frac{1}{2} 
\nabla_{a}\phi\nabla^{a}\phi -V\left(\phi
\right)\right] \,,
\end{eqnarray}
where  $\kappa \equiv 8\pi G$ and  
\begin{eqnarray}
V\left(\phi\right)=V_0 \, \mbox{e}^{-\lambda \phi} \,,
\end{eqnarray}
and $V_0$ and $ \lambda$ are two positive constants (this potential 
has been investigated in great detail in cosmology). The 
coupled Einstein-Klein-Gordon equations simplify to 
\begin{eqnarray}
&& R_{ab}=8\pi \left( \nabla_a\phi  \nabla_b \phi +g_{ab}V 
\right) \,, \label{FonarevFieldEq1}\\
&&\nonumber\\
&&\Box \phi-\frac{d
V}{d\phi}=0 \,.
\end{eqnarray}
The spherically symmetric  Fonarev line element and scalar field are 
\begin{eqnarray}
\label{eq:fonarev}
ds^2 & = & a^2 \left(\eta\right) \left[ -f^2\left(r\right) 
d\eta^2+\frac{dr^2}{f^2\left(r\right)}  
+S^2 \left(r\right) d\Omega^2_{(2)} \right] \,,\\
&&\nonumber \\
\phi \left( \eta, r \right) &=&
 \frac{1}{\sqrt{\lambda^2+2}}\ln
\left(1-\frac{2w}{r}\right) 
+\lambda\ln a
+ \frac{1}{\lambda} \ln \left[ 
\frac{V_0\left(\lambda^2-2\right)^2 }{2A_0^2 
\left(6-\lambda^2\right)} \right] 
\,,\nonumber\\
&&
\end{eqnarray}
where
\begin{eqnarray}
f(r) &=&\left(1-\frac{2w}{r}\right)^{\frac{\alpha}{2}} \,,\ 
\ \ \
\alpha =\frac{\lambda}{\sqrt{\lambda^2+2}} \,, \\
S(r)  & = & r 
\left(1-\frac{2w}{r}\right)^{\frac{1-\alpha}{2} 
} \, , \ \ \ a(\eta) =A_0 |\eta|^{\frac{2}{\lambda^2-2}} 
\,,
\end{eqnarray}
with $w$ and $ A_0$ constants and $\eta$ is the conformal time. 
For simplicity we choose 
$A_0=1$.  When $w=0$ the  metric (\ref{eq:fonarev})  
reduces to a spatially flat FLRW one while, when $a \equiv 
1$ and  $\alpha=1$, it degenerates into the Schwarzschild  
solution (however, the value $\alpha=1$ is not possible  
if $\alpha=\frac{\lambda}{\sqrt{\lambda^2+2}}$). The line 
element becomes asymptotically that of spatially flat FLRW space as 
$r\rightarrow +\infty$. The Husain-Martinez-Nu\~nez class of solutions 
(\ref{HMNconformaltime}) is recovered by setting  
$\lambda=\pm \sqrt{6}$ and $V_0=0$. See Refs.~\cite{fo:1995, HidekiFonarev} 
for the corresponding conformal diagrams.

\subsubsection{A generalized Fonarev solution}

A generalized Fonarev solution corresponding to a 
dynamical  phantom scalar field solution of GR 
is known \cite{GaoChenFaraoniShen08}. It is obtained from the 
Fonarev solution via the transformation
\begin{eqnarray} 
\phi\rightarrow i\phi \,, \ \ \ \ \lambda\rightarrow
-i\lambda \,.
\end{eqnarray}
The corresponding action is
\begin{eqnarray}
S=\frac{1}{2\kappa} \int d^4x\sqrt{-g}\left[ {R^a}_a  
+\frac{1}{2}\nabla_a 
\phi\nabla^a\phi- V\left(\phi
\right)\right] 
\end{eqnarray}
and it contains a phantom field endowed with the 
``wrong'' sign of the kinetic term.   
The generalized Fonarev line element representing a dynamical 
black hole immersed in a phantom FLRW background is 
\begin{eqnarray}
ds^2&=& a^2 \left(\eta\right) 
\left[ -f^2\left(r\right) d\eta^2 
+\frac{dr^2 }{f\left(r\right)^2}  
+S^2 \left(r\right) d\Omega^2_{(2)}\right] \,,\\
&&\nonumber \\
\phi \left( \eta, r \right)&=&\frac{1}{\lambda} 
\ln \left[ \frac{V_0\left(\lambda^2 
+2\right)^2}{2\left(\lambda^2+6\right)} \right] -\lambda\ln
a  -\frac{1}{\sqrt{\lambda^2-2}}\ln
\left(1-\frac{2w}{r}\right) \,,\nonumber\\
&&
\end{eqnarray}
where
\begin{eqnarray}
f(r) &=&\left(1-\frac{2w}{r}\right)^{\alpha/2} 
\,,\ \ \ \
\alpha=-\frac{\lambda}{\sqrt{\lambda^2-2}} \,,  
\label{xinoceros}\\
S(r) & = & r 
\left(1-\frac{2w}{r}\right)^{\frac{1-\alpha}{2}} 
\,, \ 
\ \
a(\eta) =\eta^{-\frac{2}{\lambda^2+2}}  \,.
\end{eqnarray}
Assuming that $\lambda>\sqrt{2}$, it is of interest to   
understand the physical meaning of the constant $w$. 
When $\lambda \gg \sqrt{2}$ it is  $a \approx 1 $ and 
$ \alpha \approx -1$ and the metric approximates to 
\be
\label{eqq:metric}
ds^2 \approx  -\left(1-\frac{2w}{r}\right)^{-1} 
d\eta^2+\left(1-\frac{2w}{r}\right)dr^2 
+r^2\left(1-\frac{2w}{r}\right)^{2}d\Omega^2_{(2)} \,.
\ee
The coordinate transformation \cite{GaoChenFaraoniShen08}
\begin{eqnarray}
\label{eq:coordi}y=r\left(1-\frac{2w}{r}\right) \,,
\end{eqnarray}
transforms the line element  (\ref{eqq:metric}) into  
\be
\label{eq:metricsch}
ds^2 = -\left(1+\frac{2w}{y}\right)d\eta^2+
\left(1+\frac{2w}{y}\right)^{-1}dy^2 +y^{2}d\Omega^2_{(2)} \,;
\ee
this is the Schwarzschild spacetime with 
mass $-w$. The  parameter $w$ 
corresponds to the  negative of the mass in this limit and 
from now on we will 
use $-M$ instead of  $w$.

Let us locate the apparent horizons  as the parameters $M$ and 
$\alpha$ vary. This phantom black hole solution can be 
cast in the form
\begin{eqnarray}
ds^2&=&\frac{1}{\eta^{\frac{2\alpha^2-2}{2\alpha^2-1}}} 
\left[-\left(1+\frac{2M}{r}\right)^{\alpha}d\eta^2 
+\left(1+\frac{2M}{r}\right)^{-\alpha}dr^2 
\right. \nonumber\\
&&\nonumber\\
&\, & \left.+r^2\left( 
1+\frac{2M}{r}\right)^{1+\alpha}d\Omega^2_{(2)}\right] \,;
\end{eqnarray}
the replacement of the conformal time $\eta$ with the 
comoving time $t$ leads to
\begin{eqnarray}
\label{eq:cosmic time}
ds^2&=&-\left(1+\frac{2M}{r}\right)
^{\alpha}dt^2 \nonumber\\
&&\nonumber\\
&\, & +a^2 \left(t\right) \left[ 
\left(1+\frac{2M}{r}\right)^{-\alpha}dr^2
+r^2\left(1+\frac{2M}{r}\right)^{1 
+\alpha}d\Omega^2_{(2)}\right] \,,\\
&&\nonumber \\
a\left(t\right)&=& \left(t_0-t\right)^{ 
-\frac{2\left( \alpha^2-1\right)}{\alpha^2}} \,,\nonumber\\
&& 
\end{eqnarray}
where the integration constant $t_0$ marks the time
of the Big 
Rip and it is $\alpha<-1$ since $\lambda>\sqrt{2}$.  
The exponent $\alpha$ is determined by
the slope of the potential according to eq. 
(\ref{xinoceros}).  
When $M=0$ the spacetime~(\ref{eq:cosmic time}) 
reduces to a phantom-dominated FLRW cosmos. By setting, for 
simplicity,  $\alpha=-3$ or  $\lambda={3}/{2}$, 
the line element (\ref{eq:cosmic time}) reduces to 
\begin{eqnarray} 
ds^2&=&-\left(1+\frac{2M}{r}\right)
^{-3}dt^2 \nonumber\\
&&\nonumber\\
&\, & +a^2 \left(t\right) \left[\left( 
1+\frac{2M}{r}\right)^{3}dr^2
+r^2\left(1+ 
\frac{2M}{r}\right)^{-2}d\Omega^2_{(2)} \right] 
\,,\nonumber\\
&&\nonumber \\
a\left(t\right) 
&=&\left(t_0-t\right)^{-16/9} \,.
\end{eqnarray}
In terms of the areal radius $R=ar\left( 1+2M/r 
\right)^{-1}$, the equation locating the apparent horizons 
is  
\be 
1+\frac{8Ma}{R}\left( 
1+\sqrt{1+\frac{8Ma}{R}} \, \right)^{-1}
-\frac{HR}{32}\left( 
1+\sqrt{1+\frac{8Ma}{R}} \, \right)^5=0 \,,
\ee
where  $H\equiv \dot{a}/a$ is the Hubble parameter of the 
background.  Further setting $x\equiv
1+\sqrt{1+\frac{8Ma}{R}}$ yields 
\begin{equation}
 \label{eq:ah} aMHx^4-4x^2+12x-8=0 \, .
\end{equation}
This quartic equation has only two real positive roots 
corresponding to a cosmological apparent horizon $R_c$ and 
a black hole apparent horizon $R_b$ 
\cite{GaoChenFaraoniShen08}. The qualitative 
behaviour of the apparent horizons is the same as that of 
the McVittie and generalized McVittie classes of solutions with a phantom
FLRW substratum: 
a black hole apparent horizon inflates while a cosmological  
apparent horizon shrinks. At a critical time these two 
apparent horizons meet and disappear leaving behind a 
naked singularity \cite{GaoChenFaraoniShen08}.

\subsection{The Swiss-Cheese model}

In 1945, apparently unaware of McVittie's work from a decade earlier, 
Einstein and Straus \cite{EinsteinStraus, EinsteinStraus2} constructed 
the solution of GR now called ``Einstein-Straus vacuole'' or 
``Swiss-cheese model'' by pasting a Schwarzschild-like region of 
spacetime onto a dust-dominated FLRW universe across a timelike 
hypersurface (for reviews see \cite{Krasinskibook, CarreraGiuliniRMD10, 
MarsMenaVera13}, which we partially follow here). There is a black hole 
event horizon in this spacetime and the usual energy conditions are 
satisfied.

Let the interior Schwarzschild region be denoted with 
${\cal M}^{-}$ and the exterior FLRW region with ${\cal 
M}^+$ and let $\Sigma$ be a spacelike 2-sphere of constant 
comoving 
radius $r_{\Sigma}$. The coordinate charts covering 
$\Sigma$ are the FLRW $ \left\{ t, \theta, \varphi 
\right\}$ and the Schwarzschild chart
$\left\{ T(t), \theta, \varphi \right\}$. The metric in 
the two regions is given by
\begin{eqnarray}
ds^2_{(-)} & = & -\left( 1-\frac{2m}{R} \right)dT^2 
+\frac{dR^2}{1-\frac{2m}{R}} +R^2 d\Omega_{(2)}^2 \,,
\nonumber\\
&&\nonumber\\
ds^2_{(+)} & = & -dt^2  +a^2(t) \left( \frac{dr^2}{1-kr^2}  
 +r^2 d\Omega_{(2)}^2 \right) \,.
\end{eqnarray}

Choose on $\Sigma$ the triad of orthonormal vectors
\be
\left\{ e^{\alpha}_{(t)}, e^{\alpha}_{( \theta)}, 
e^{\alpha}_{( \varphi)} \right\}= 
\left\{ \delta^{\alpha}_t, \frac{ 
\delta^{\alpha}_{\theta}}{ar} , 
\frac{ \delta^{\alpha}_{\varphi}}{ar\sin\theta} \right\} 
\,,
\ee
where $\alpha, \beta=r, \theta, \varphi$. 
The equation of $\Sigma$ is $\Phi(r) \equiv  r-r_{\Sigma}=0$ 
and the gradient of $\Phi$ is $N_a \equiv \nabla_a \Phi= \delta _{a 
r}$, with norm  squared $N_a N^a= g^{rr}=\frac{1-kr^2}{a^2} 
$. The unit normal to $\Sigma$, therefore, has components
\be
n_{\mu}=\frac{ N_{\mu}}{\sqrt{N_{\nu}N^{\nu}}}=\left( 0, 
\frac{a}{\sqrt{1-kr^2}}, 0, 0 \right) \,.
\ee
The extrinsic curvature of $\Sigma$ is given by the usual 
formula
\be
K_{\alpha\beta}= e^{(a)}_{\alpha}e^{(b)}_{\beta}  
\nabla_an_b
\ee
which is used to compute $K_{\alpha\beta}$ in ${\cal M}^{-}$ and 
${\cal M}^{+}$.  
The continuity of the first and second 
fundamental forms on $\Sigma$ requires 
\begin{eqnarray}
&& R_{\Sigma}(t)=a(t) r_{\Sigma} \,,\\
&&\nonumber\\
&& \left( 1-\frac{2m}{R_{\Sigma}} \right) \left( 
\frac{dT}{dt} 
\right)^2 - \left( \frac{dR}{dt} \right)^2 
\frac{1}{1-\frac{2m}{R_{\Sigma}} }=1 \,,\\
\end{eqnarray}
the combination of which yields
\be
\frac{dT}{dt}=\left( 1-\frac{2m}{R_{\Sigma}} \right)^{-1} 
\sqrt{ 1-\frac{2m}{R_{\Sigma}} +H^2 R^2_{\Sigma} } \,.
\ee
Using the Hamiltonian constraint of FLRW space 
$H^2=\frac{8\pi}{3} \, \rho -\frac{k}{a^2}$, one obtains
\be
1-\frac{2m}{R_{\Sigma}} +H^2 R^2_{\Sigma} = 1-kr_{\Sigma}^2 
+ \left( \frac{8\pi}{3} \, \rho R_{\Sigma}^2 -\frac{2m}{ar_{\Sigma}} 
\right) \,. \label{mmmmmatch}
\ee
In the absence of surface distributions of mass-energy 
on $\Sigma$, the stress-energy tensor of the matter 
source of this solution of the Einstein equations must also 
be continuous across $\Sigma$. Since the interior is 
vacuum, the pressure on the outside is forced to vanish, 
$P^{(+)}=P^{(-)}=0$, which implies that only a 
dust-dominated FLRW background can match the 
Schwarzschild solution. Moreover, the energy density must be 
continuous at $\Sigma$, implying that  
\be
\rho_{\Sigma} =  \frac{m}{ \frac{4\pi}{3} \, R^3_{\Sigma}} \,,
\ee
which means that the mass of the black hole inside the 
vacuole must equal  the mass that a sphere of volume 
$4\pi R_{\Sigma}^3/3$ would have in the FLRW 
background (note that this  
volume is not the proper volume of such a sphere unless the 
FLRW curvature index $k$ vanishes). This condition yields $ 
\frac{8\pi }{3} \, \rho R_{\Sigma}^2  
=  \frac{2m}{ar_{\Sigma}} $. Eq. (\ref{mmmmmatch}) then 
reduces to 
\be
1-\frac{2m}{R_{\Sigma}} +H^2 R^2_{\Sigma} = 1-kr_{\Sigma}^2 
\,.
\ee
The continuity of the matter distribution across $\Sigma$ can be seen as the 
continuity of the Misner-Sharp-Hernandez mass  $ M^{(+)}_{MSH}= M^{(-)}_{MSH}$ 
\cite{Eisenstaedt77} (see \cite{CarreraGiuliniRMD10, Krasinskibook} for a 
detailed discussion).

The Einstein-Straus model has no accretion onto the central 
inhomogeneity. The interior Schwarzschild region is 
shielded from the cosmological expansion (and also the exterior FLRW region sees 
no effect from the central hole) and is static and, 
because of this fact, the Swiss-cheese model is often used as 
supporting evidence that the cosmological expansion does 
not affect local systems. However, the boundary of the 
vacuole is expanding and perfectly comoving; if the 
vacuole is regarded as the ``local object'' (instead of the 
black hole in it, which is insulated by a vacuum region), this argument 
fails. 
The Einstein-Straus vacuole has  few drawbacks: 
it is  unable to describe  the  Solar System   
\cite{Bonnoratom,  Krasinskibook} and  is unstable with 
respect to  non-spherical perturbations 
\cite{SenovillaVera97, Mars98,  NolanPRD98, 
MenaTavakolVera02} 
and to perturbations of the matching 
condition $ M_{MSH}^{(+)}=  M_{MSH}^{(-)}$ 
\cite{Krasinskibook}. 

The Einstein-Straus vacuole was generalized  
to include a cosmological constant, 
obtaining a Schwarschild-(anti-)de Sitter  instead 
of Schwarz\-schild  interior \cite{Balbinotetal88},  or 
to include a fluid with pressure in the 
interior region \cite{BonaStela87}. Also the generalization 
obtained by matching a Schwarschild interior with an  
inhomogeneous Lema\^itre-Tolman-Bondi exterior has 
been studied \cite{Bonnor00}.   The Hawking 
radiation emitted by the Einstein-Straus black hole has 
been studied in \cite{Saida02, SaidaHaradaMaeda07}. It is 
found that such a black hole in an expanding
universe is excited to a non-equilibrium state and 
emits with stronger intensity than a thermal one.

\subsection{Other GR solutions}

There are several other analytical solutions of the Einstein equations 
describing central inhomogeneities in FLRW backgrounds. While one has to be careful  
as many of them do not have reasonable matter sources, 
they are of some interest.  They 
cannot be included here due to space limitations (for a more general and 
rigorous treatment of inhomogeneous cosmologies see the book by 
Krasi\'nski \cite{Krasinskibook}). They include, among others, the well 
known Lema\^itre-Tolman-Bondi and Szekeres solutions ({\em e.g.}, 
\cite{Krasinskibook, Goncalves01, BoothBritsGonzalezVDB, 
 LTBwithGao, F1, F2, F3, F4, F5, KrasinskiHellaby04}), the Oppenheimer-Snyder solution 
\cite{OppenheimerSnyder, BenDov05, BoothBritsGonzalezVDB}, members of the large Barnes 
family \cite{Barnes73}, the solutions of Dyer, McClure, and collaborators 
\cite{McClureDyer06CQG, McClureDyerGRG, McClureAndersonBardahl07, 
McClureAndersonBardahl08, McClurethesis}, the Roberts solution with a scalar field 
\cite{Roberts89, Burko97}, 
Patel and Trivedi's \cite{PatelTrivedi82} and  Vaidya's Kerr-FLRW solutions  
\cite{Vaidya77}, Balbinot's evaporating black hole \cite{Balbinot}, 
and other solutions can be obtained from the previous ones with 
cut-and-paste techniques \cite{Nayak, Cox03}, possibly to excise regions in which the energy 
conditions are violated. Also asymptotically flat metrics describing 
transient and time-dependent horizons have been studied \cite{Lindesay1, Lindesay2, 
Lindesay3, Lindesay4, Lindesay5, Lindesay6, Lindesay7, Lindesay8, Adleretal05} 
and other solutions 
are of potential interest \cite{Cardosoetal, KastorTraschen93, Brilletal94, 
KoberleinMallett94, Husain96, 
DawoodGosh04, ConboyLake07, KyoHaradaMaeda08, Meissner09, 
GibbonsMaeda10, Maeda12, Culetu12}.

\section{Some cosmological black holes and naked singularities in alternative
 gravity}

Few solutions of theories of gravity alternative to GR and representing 
cosmological black holes at least part of the time are known, most of them in 
scalar-tensor gravity. Here we review a few. The simplest scalar-tensor theory, 
Brans-Dicke gravity, is described by the action \cite{BransDicke61} 
\be 
S_{BD}=\int d^4x \, \sqrt{-g} \left[ \phi {R^c}_c -\frac{\omega}{\phi} \, 
g^{ab} \nabla_{a}\phi \nabla_{b}\phi +2\kappa \,{\cal L}^{(m)} \right] 
\,, 
\ee 
where ${\cal L}^{(m)}$ is the matter 
Lagrangian, $\phi$ is the Brans-Dicke scalar field (roughly speaking, the 
inverse of the effective gravitational coupling strength), and $\omega$ is a 
parameter (``Brans-Dicke coupling''). In more general scalar-tensor theories 
\cite{Bergmann68, Wagoner70, Nordtvedt70}, 
the Brans-Dicke coupling is promoted to a function of $\phi$,
 $\omega=\omega(\phi)$.

\subsection{The conformal cousin of the Husain-Martinez-Nu\~nez solution}

A solution of Brans-Dicke gravity was generated, but not interpreted,
 by Clifton, Mota, and Barrow \cite{CMB05} 
by conformally transforming the Husain-Martinez-Nu\~nez solution, $ 
g_{\mu\nu}^{(HMN)} \longrightarrow \Omega^2\, g_{\mu\nu}^{(HMN)}=\phi 
g_{\mu\nu}^{(HMN)} $ with $ \phi \longrightarrow 
\tilde{\phi}=\sqrt{\frac{2\omega+3}{16\pi}} \, \ln \phi $. This is the inverse 
of the usual transformation from the Jordan frame to the Einstein frame which 
turns gravity with a scalar field non-minimally coupled to the Ricci scalar 
into GR with a scalar field with canonical kinetic energy but 
non-minimally coupled to matter. The Clifton-Mota-Barrow solution is 
\begin{eqnarray} 
ds^2 &=& -A^{\alpha \left( 
1-\frac{1}{\sqrt{3}\, \beta}\right)} (r) \, dt^2 \\ &\,& +A^{-\alpha \left( 
1+\frac{1}{\sqrt{3}\, \beta}\right)} (r) \, t^{\frac{2 \left( 
\beta-\sqrt{3}\right)}{3\beta - \sqrt{3}} } \left[ dr^2 + r^2 A(r) 
d\Omega_{(2)}^2 \right] \,,\\ 
&&\nonumber\\
\phi( t,r) &=& A^{\frac{\pm 1}{2\beta}} (r)\, 
t^{\frac{2}{\sqrt{3}\, \beta -1}} \,, 
\end{eqnarray} 
where 
\be 
 A(r) = 1-\frac{2C}{r} 
\,,\;\;\; \beta =\sqrt{2\omega+3} \,,\;\; \omega>-3/2 \,, \;\;
\alpha=\pm \sqrt{3}/2 \,.
\ee 
There are singularities at 
$r=2C$ and at $t=0$ (here it must be 
$2C<r<+\infty$ and $t>0$). The scale factor 
of the spatially flat FLRW background is 
\be
 a(t)= t^{\frac{\beta-\sqrt{3}}{3\beta-\sqrt{3}} }\equiv t^{\gamma} \,. 
\ee 
The solution was interpreted in Ref.~\cite{CMBandres}.
 We rewrite  the 2-parameter line element as 
\be
 ds^2=-A^{\sigma}(r) \, dt^2 +A^{\Theta}(r) \, 
a^2(t)dr^2 +R^2(t,r)d\Omega_{(2)}^2 \,, \label{lineelementq} 
\ee 
where 
\be 
\sigma = \alpha \left( 1-\frac{1}{\sqrt{3}\, \beta} \right) \,, \;\;\;\; \Theta 
= -\alpha \left( 1+\frac{1}{\sqrt{3}\, \beta} \right) \,, 
\ee 
and 
\be 
R(t,r)=A^{\frac{\Theta+1}{2}}(r) \, a(t) \,r  
\ee 
is the areal radius. 
It is useful to study the area of the 2-spheres of symmetry: we have $
 \partial R/\partial r
 =  a(t)A^{\frac{\Theta-1}{2}}(r) 
\left( 1- r_0/r \right) $ where $ 
r_0=(1-\Theta)C$ or
\be
R_0(t)=\left( \frac{\Theta+1}{\Theta-1} 
\right)^{\frac{\Theta+1}{2}}  (1-\Theta) a(t) \, C \,.
\ee
The critical value $r_0$ lies in the physical spacetime 
region  $r_0>2C$ if $\Theta<-1$. $R$ has the limit 
\be
R(t, r)=\frac{ r \, a(t) }{ \left(1-\frac{2C}{r} 
\right)^{|\frac{\Theta+1}{2}|}} \rightarrow +\infty \;\;
\mbox{as } r\rightarrow 2C^{+}
\ee
For $\Theta<-1$, the areal radius $R(r)$ has  a minimum at 
$r_0$, the area $4\pi R^2$ of the 2-spheres of symmetry is minimum 
there, and there is a wormhole throat joining two spacetime regions. 
Since 
\be
\Theta= \mp \frac{\sqrt{3}}{2} \left( 1+ \frac{1}{\sqrt{3} 
\sqrt{2\omega+3} } \right) 
\ee
for $\alpha=\pm \sqrt{3}/2$, the condition $\Theta<-1$ 
requires $\alpha =+\sqrt{3}/2$ (this is a necessary but not  
sufficient condition for the throat to exist). The 
sufficient condition $\Theta<-1$ constrains the Brans-Dicke  
parameter as \cite{CMBandres}
\be
\omega < \frac{1}{2} \left[ \frac{1}{ \left( 2-\sqrt{3} 
\right)^2}-3 \right] \simeq 5.46 \equiv \omega_0 \,.
\ee
For   $-3/2< \omega <\omega_0 $ the  solution can be interpreted as  
a cosmological Brans-Dicke wormhole.  The region $2C<r<r_0$ is not a FLRW 
region and the scalar field  is finite and non-zero at $r_0$: 
\be
\phi\left(t, r_0 \right)= t^{ \frac{2}{\sqrt[]{3}\beta -1 
} } \left(  \frac{\Theta+1}{\Theta-1} \right)^{\frac{\pm 
1}{2\beta} } 
\,.
\ee
The wormhole throat is exactly comoving with the cosmic  
substratum, which is relevant for the problem of cosmic expansion versus  
local systems \cite{CarreraGiuliniRMD10} and disappears in the limit 
$C\rightarrow 0$.  

Let us study the existence and location of the apparent horizons of this 
spacetime. The relation between differentials
\be
dr=\frac{ dR-A^{\frac{\Theta+1}{2}}(r) \, \dot{a}(t)rdt}{
A^{\frac{\Theta-1}{2}}a(t) \, \frac{C( \Theta+1)}{r} 
+A^{\frac{\Theta+1}{2}}(r) \, a(t)}\,,
\ee
turns the line element into
\be
ds^2 = -A^{\sigma}dt^2 +\left[  
\frac{dR^2-2A^{\frac{\Theta+1}{2}} r\dot{a} \, dtdR 
+A^{\frac{\Theta+1}{2}} 
r^2 \dot{a}^2 dt^2}{ D_1(r)} \right] 
 +R^2 d\Omega_{(2)}^2 \,,
\ee
where
\be
D_1(r)=A(r) \left[ 1+\frac{C(\Theta+1)}{r \,A(r)} \right]^2 
\,.
\ee
Starightforward manipulations yield  
\be
ds^2 = - \frac{ \left( D_1A^{\sigma} -H^2R^2\right)}{D_1} 
\, dt^2 -\frac{2HR}{D_1}\, dtdR +\frac{dR^2}{D_1} 
 + R^2  d\Omega_{(2)}^2 
\ee
where $H \equiv \dot{a}/a$.  The inverse metric 
in coordinates $\left( t, R, \theta, \varphi \right)$  is
\be
\left( g^{\mu\nu} \right)=\left(
\begin{array}{cccc}
-\frac{1}{A^{\sigma}} & -\frac{HR}{A^{\sigma}} & 0 & 0 \\
&&&\\
-\frac{HR}{A^{\sigma}} &\frac{\left( D_1 
A^{\sigma}-H^2R^2\right)}{A^{\sigma}} & 0 & 0  \\
&&&\\
0 & 0 & R^{-2} & 0 \\
&&&\\
0 & 0 & 0 &R^{-2}\sin^{-2}\theta 
\end{array} \right).
\ee
The apparent horizons are located by the roots of     
$g^{RR}=0 $, or
\be\label{struzzob}
D_1 (r) A(r)= H^2(t) R^2(t,r) 
\ee
There are solutions which describe apparent horizons with the ``S-curve'' phenomenology 
of the Husain-Martinez-Nu\~nez solution of GR. Eq.~(\ref{struzzob}) is 
satisfied also if the right hand side is time-independent, $H=\gamma/t =0$, $\gamma=0$, 
$\beta=\sqrt{3}$, $\omega=0$, which produces a static Brans-Dicke  solution 
describing an inhomogeneity in a Minkowski background. 

Other cases include: a) $\omega \geq \omega_0$ and b) $\alpha= - 
\sqrt{3}/2$. 
In both cases there are no wormhole throats and no apparent horizons 
and the Clifton-Mota-Barrow spacetime contains a naked singularity.

For $\alpha=-\sqrt{3}/2$ it is $ \Theta =\frac{\sqrt{3}}{2}  
\left( 1+\frac{1}{\sqrt{3}\, \beta} \right)>0$ and 
\be 
R(t, r)=r \left( 1-\frac{2C}{r} 
\right)^{\frac{|\Theta+1|}{2}} 
a(t)\rightarrow 0 \;\;\; \mbox{as }  r\rightarrow 2C^{+}\,.
\ee
Since $r_0<2C$, the areal radius $R(r)$ always increases for  
$2C<r<+\infty$ and there is a naked singularity at $R=0$.

Let us consider now the special case $\omega=0$: this value  of the 
Brans-Dicke  coupling (corresponding to $ \beta=\sqrt{3}$ and $\gamma=0$) 
produces the static 
metric 
\be\label{StaticMetric} 
ds^2=-A^{\frac{2\alpha}{3}}(r) dt^2 
+\frac{dr^2}{A^{\frac{4\alpha}{3}}(r)} 
+\frac{r^2}{A^{\frac{4\alpha}{3}-1}(r)}\, d\Omega_{(2)}^2 
\ee 
and the scalar field 
\be\label{StaticScalar} 
\phi(t,r)=A^{ \frac{\pm 1}{2\sqrt{3}}}(r) t \,, 
\ee 
which is time-dependent even though the metric is static (this is 
analogous to another Brans-Dicke solution 
\cite{VFVitaglianoSotiriouLiberati12}). 
The metric (\ref{StaticMetric}) is a Campanelli-Lousto metric 
\cite{CampanelliLousto1, CampanelliLousto2}. 
The general Campanelli-Lousto solution of Brans-Dicke theory 
has the form 
\begin{eqnarray} 
ds^2 &=& 
-A^{b+1}(r)dt^2+\frac{dr^2}{A^{a+1}(r)}+ \frac{r^2 d\Omega_{(2)}^2}{A^a(r)} 
\,,\\ 
&&\nonumber\\ 
\phi(r) & = & \phi_0 A^{\frac{a-b}{2}}(r) \,, 
\end{eqnarray} 
with $\phi_0>0, a, b$ constants, and Brans-Dicke parameter 
\be 
\omega(a,b)= -2 \left( a^2+b^2-ab+a+b\right) \left( a-b \right)^{-2} \,. 
\ee 
In our case, setting $ \left( a, b \right)= \left( \frac{4\alpha}{3} -1, 
\frac{2\alpha}{3} -1 \right) $ reproduces the Campanelli-Lousto metric.  
Then, $\omega \left( 
\frac{4\alpha}{3}-1, \frac{2\alpha}{3}-1 \right)=0$ for $\alpha =\pm 
\sqrt{3}/2$. The scalar field differs from the Campanelli-Lousto one by the linear 
dependence on $t$, hence the static limit of the Clifton-Mota-Barrow solution is a 
trivial generalization of a Campanelli-Lousto solution.  
The nature of the Campanelli-Lousto spacetime 
depends on the sign of $a$ which, for us, corresponds to the choice $\alpha=\pm 
\sqrt{3}/2$ \cite{VanzoZerbiniVF12}.  For $a \geq 0$ 
(which corresponds to $ \alpha=+\sqrt{3}/2, a \simeq 0.1547$, and $ \Theta 
=-\frac{4\alpha}{3} \simeq -1.1547 <-1$) the Campanelli-Lousto 
spacetime contains a wormhole throat which coincides with an apparent horizon 
at $r_0=2C \left( \frac{1-\Theta}{2}\right)>2C$ \cite{VanzoZerbiniVF12}.

For $a<0$ (or $  \alpha=-\sqrt{3}/2\,, a\simeq 
-2.1547$, and $ \Theta \simeq 1.1547>0$), there are no apparent horizons and 
the spacetime contains a naked singularity. An explanation of why the 
Husain-Martinez-Nu\~nez spacetime describes a black hole but the  conformally 
related Clifton-Mota-Barrow  one does not is given in \cite{CMBandres}.

\subsection{The Brans-Dicke solutions of Clifton, Mota, and Barrow}

The next class of solutions of the Brans--Dicke field equations that we 
consider is that of Clifton, Mota, and Barrow given by the time-dependent 
and spherically symmetric line element \cite{CMB05}
\be\label{CMBmetric}
ds^2=-e^{\nu (\varrho)}dt^2+a^2(t) e^{\mu (\varrho)}(d\varrho^2+\varrho^2d\Omega^2) \,,
\ee
where
\begin{eqnarray} 
e^{\nu (\varrho)} & = &  
\left(\frac{1-\frac{m}{2\alpha \varrho}}{1+\frac{m}{2 \alpha \varrho}}
\right)^{2\alpha 
}\equiv A^{2\alpha} \,,\\
&&\nonumber\\
e^{\mu (\varrho)} & = & \left(1+\frac{m}{2\alpha \varrho}\right)^{4} 
A^{\frac{2}{\alpha}( \alpha-1)(\alpha +2)} \,, \\
&&\nonumber\\
\label{abeta}
a(t) & = & a_0\left(\frac{t}{t_0}\right)^{\frac{ 
2\omega (2-\gamma)+2}{3\omega \gamma(2-\gamma)+4}}\equiv 
a_{\ast}t^{\beta} \,,\\
&&\nonumber\\
\label{scalart}
\phi(t, \varrho) &= & \phi_0\left(\frac{t}{t_0}\right)^{\frac{2(4-3\gamma)}{ 
3\omega \gamma(2-\gamma)+4}}A^{-\frac{2}{\alpha }(\alpha^2-1)} \,,\\
&&\nonumber\\
\alpha & = & \sqrt{ \frac{ 2( \omega +2 )}{2\omega +3} } 
\,,\label{7}\\
&&\nonumber\\
\rho^{(m)}(t, \varrho) & = & \rho_0^{(m)} \left( \frac{ a_0}{a(t)} 
\right)^{3\gamma} A^{-2\alpha} \,, \label{density}
\end{eqnarray}
The matter source is a perfect fluid with 
energy density and pressure $\rho^{(m)}$ and  $P^{(m)}$ and equation of 
state $P^{(m)}=\left( \gamma -1 \right) \rho^{(m)}$ with  
$\gamma=$const. \cite{CMB05}.  $m$ is a mass parameter and $\alpha, 
\phi_0, a_0$, $\rho^{(m)}_0$ and 
$t_0$ are positive constants ($\phi_0$, $\rho^{(m)}_0$ and 
$t_0$ are related).  $\varrho$ is the isotropic 
radius  related to the  areal radius Schwarzschild 
$\tilde{r}$ by 
\be\label{isotropicradius}
\tilde{r} \equiv \varrho \left(1+\frac{m}{2\alpha \varrho}\right)^{2}\,,
\ee
and 
\be
d\tilde{r}=\left(1-\frac{m^2}{4\alpha^2\varrho^2}\right)d\varrho \,.
\ee
$\alpha $ is real for $\omega <-2$ and for $ \omega > 
-3/2$ but, for brevity, we require that $\omega_0>-3/2$ and $ \beta \geq 
0$. 
The line element ~(\ref{CMBmetric}) reduces to the spatially flat FLRW one  
when $m\rightarrow 0$.  For $\gamma \neq 2$, setting 
$\omega =\left( \gamma -2 \right)^{-1}$ produces $\beta =0$ and 
the spacetime becomes static (but the scalar still depends on time). 
The values $\gamma=2$ and $\gamma=4/3$ of the $\gamma$-parameter yield 
$\beta =1/2$ and $a(t)\sim \sqrt{t}$ independent of the parameter  
$\omega $.

To interpret the solution physically, the apparent horizons were studied in 
Ref.~\cite{VFVitaglianoSotiriouLiberati12}.  Again, the analysis proceeds by 
rewriting the line element using the areal radius 
\be\label{arealradius} 
r = a (t) \varrho \left( 1+\frac{m}{2\alpha \varrho} \right)^{2} 
A^{\frac{1}{\alpha}(\alpha -1)(\alpha+2)}  =  a(t) 
\tilde{r} A^{\frac{1}{\alpha}( \alpha -1)( \alpha +2)} 
\ee
and solving the 
equation $g^{RR}=0$ numerically.  It was found that, according to the parameter 
values, several kinds of behaviours are possible. The ``S-curve'' familiar from 
the Husain-Martinez-Nu\~nez solution of GR is reproduced in a certain region of 
the parameter space, but different behaviours appear for other combinations of 
the parameters \cite{VFVitaglianoSotiriouLiberati12}. In certain regions of the 
parameter space, the Clifton-Mota-Barrow spacetime contains a naked 
singularity created with the universe and the metric and scalar field cannot 
be obtained as regular developments of Cauchy data. Later, this singularity  
is covered by 
black hole apparent horizons. In other regions of the parameter space, pairs of 
black hole and cosmological apparent horizons appear and bifurcate, or merge and 
disappear, as in the Husain-Martinez-Nu\~nez solution of GR  and in 
the Clifton solution of  $f\left({R^c}_c\right)$ gravity 
\cite{CliftonCQG06, myClifton}. 
Due to the larger parameter space involved with respect to what seen thus 
far, the Clifton-Mota-Barrow class of spacetimes exhibits the most 
varied and  rich phenomenology of  apparent horizons seen here (including some new one  
reported in \cite{VFVitaglianoSotiriouLiberati12}).

\subsection{Clifton's  solution of $f\left({R^c}_c \right)$ gravity}

Metric $f\left({R^c}_c\right)$ gravity is described by the action
\begin{equation}
S=\frac{1}{2\kappa} \int d^4x \sqrt{-g}\, f({R^c}_c) +S^{(matter)} \,,
\end{equation}
where $f({R^c}_c)$ is a non-linear function of its argument and  
$S^{(matter)} $ is the matter part of the action. As usual, ${R^c}_c$ 
denotes  the Ricci scalar of the metric $g_{ab}$.  

The Jebsen-Birkhoff theorem of GR fails in these theories, as well as in 
scalar-tensor gravity, adding to the variety of spherical solutions 
\cite{BirkhoffPRD}. Of particular interest are black holes in these higher 
order gravity theories.

A solution of vacuum $f({R^c}_c)=\left( {R^c}_c\right)^{1+\delta}$ 
gravity was found by Clifton \cite{CliftonCQG06} and is given by the line element
\begin{equation}\label{1}
ds^2=-A_2(r)dt^2+a^2(t)B_2(r)\left[ dr^2 +r^2 \left(  d\theta^2 
+\sin^2 \theta \, d\varphi^2  \right) \right] 
\end{equation}
in terms of the  isotropic radius, where 
\begin{eqnarray}
A_2(r) &=& \left( \frac{1-C_2/r}{1+C_2/r}\right)^{2/q} \,, 
\label{2} \\
&&\nonumber \\
B_2(r) &=& \left( 1+\frac{C_2}{r} \right)^{4}A_2(r)^{\, q+2\delta 
-1} \,,\label{4}\\
&&\nonumber \\
a(t) &= & t^{\frac{ \delta \left( 1+2\delta \right)}{1-\delta}} 
\,,\label{3again}\\
&&\nonumber\\
q^2 &= & 1-2\delta+4\delta^2 \,. 
\end{eqnarray}
 Since Solar System tests require    
$\delta   =\left( -1.1\pm 1.2 \right) \cdot 10^{-5}$ 
 \cite{CliftonCQG06, BarrowClifton2, BarrowClifton3} and it must be 
 $f''\left({R^c}_c\right)  \geq 0$ for local stability \cite{DolgovKawasaki, 
mattmodgrav1, mattmodgrav2, Odintsovconfirm}, we assume  
$0<\delta < 10^{-5}$. Once $\delta $ is fixed, two classes of solutions exist, 
corresponding to the sign of  $C_2 q r $.  The line 
element~(\ref{1}) becomes  FLRW  if  $C_2\rightarrow 0$. 
If $\delta \rightarrow 0$ (in which case the theory reduces to 
GR), the line element~(\ref{1}) reduces to the 
Schwarzschild one provided that 
$C_2 q r>0$. In principle both positive and negative values of $r$ are 
possible according to the sign of $C_2$, but we impose that   
$r>0, C_2>0$ and we take the positive root in the expression $q=\pm 
\sqrt{1-2\delta +4\delta^2}$. Then  
$ q\simeq 1-\delta$ as $ \delta \rightarrow 0 $.  
The Clifton solution  is conformal to the 
Fonarev  spacetime \cite{fo:1995},  which is conformally static 
\cite{HidekiFonarev}, hence it is also conformally static.  
Clifton's  solution is dynamical and represents 
a central inhomogeneity in a spatially flat FLRW 
universe in vacuum $ \left({R^c}_c \right)^{1+\delta}$ gravity.
In the fourth order field equations of  metric $f\left({R^c}_c\right)$ gravity
\begin{equation}
f'\left( {R^c}_c \right) R_{ab}-\frac{f\left({R^c}_c\right)}{2}
\, g_{ab}=\nabla_a\nabla_b 
f'\left({R^c}_c\right)-g_{ab} \Box f'\left({R^c}_c\right) 
\end{equation}
geometric terms play the role of an effective matter which 
invalidates the Jebsen-Birkhoff theorem of GR. 
Metric $f\left({R^c}_c\right)$ gravity has an equivalent representation as an 
$\omega=0$  Brans-Dicke  theory with a special potential 
for the scalar field degree of freedom $f'\left( {R^c}_c\right)$  
\cite{SotiriouFaraoni10review, DeFeliceTsujikawa10}.

The apparent horizons of the Clifton solution \cite{CliftonCQG06} 
were studied in Ref.~\cite{myClifton}. First using   
the new radial coordinate
\begin{equation} \label{6}
\tilde{r} \equiv r \left( 1+\frac{C_2}{r} \right)^2 \,,
\end{equation}
with $dr=\left( 1-\frac{C_2^2}{r^2} \right)^{-1} 
d\tilde{r} $ and then using the areal radius
\begin{equation}\label{8again}
R \equiv \frac{ a(t) \sqrt{B_2(r)} \, \tilde{r} }{\left( 
1+\frac{C_2}{r} \right)^2} =a(t) \, \tilde{r} \, A_2(r)^{ 
\frac{q+2\delta -1}{2}}  \,,
\end{equation}
the metric~(\ref{1}) is rewritten as 
\begin{equation}\label{9}
ds^2=-A_2dt^2 +a^2 A_2^{2\delta -1}d\tilde{r}^2 +R^2 
d\Omega^2_{(2)}  \,.
\end{equation}
Then the identities
\begin{equation} \label{10}
d\tilde{r}=\frac{
dR-A_2^{\frac{q+2\delta -1}{2} } \dot{a} \, \tilde{r} \, dt}{
a\left[ A_2^{\frac{q+2\delta -1}{2}} +\frac{2( q+2\delta -1)}{q} 
\frac{C_2}{\tilde{r}} A_2^{\frac{2\delta -1-q}{2}} \right]} 
\equiv 
\frac{ dR -A_2^{\frac{q+2\delta-1}{2}} \dot{a} \, \tilde{r} \, 
dt 
}{a 
A_2^{\frac{q+2\delta-1}{2}} C(r)} \,,
\end{equation}
yield 
\begin{equation}\label{11}
C(r)=1+\frac{2(q+2\delta -1)}{q} \, \frac{C_2}{\tilde{r}} \, 
A_2^{-q} = 
1+\frac{2(q+2\delta -1)}{q} \, \frac{C_2 a}{R} \, 
A_2^{\frac{2\delta -1-q}{2}} \,,
\end{equation}
and  the metric turns into the form
\begin{eqnarray}
ds^2 & = & - A_2\left[ 1-\frac{A_2^{ 2(\delta -1)} }{C^2}\, 
\dot{a}^2 \tilde{r}^2 \right] 
dt^2-\frac{2A_2^{\frac{-q+2\delta-1}{2}}}{C^2} \, 
\dot{a} \, \tilde{r} \, dtdR \nonumber\\
&&\nonumber\\
& \,  & + \frac{dR^2}{A_2^q C^2}+ R^2d\Omega^2_{(2)} \,.\label{12}
\end{eqnarray}
Let us pass now to the new time $\bar{t}$ defined by  
\begin{equation}\label{13}
d\bar{t}=\frac{1}{F(t, R)} \left[ dt +  \beta(t,R)dR 
\right] 
\end{equation}
to eliminate the time-radius cross-term, where $F(t,R)$ 
is an integrating factor \cite{myClifton}. The line element is 
rewritten as 
\begin{eqnarray}
ds^2 & = & -A_2\left[ 1-\frac{A_2^{ 2(\delta -1)} }{ 
C^2}\,  \dot{a}^2 \tilde{r}^2 \right] F^2 d\bar{t}^2 
\nonumber\\
&&\nonumber\\
 &\, & +2F  
\left\{ A_2\beta \left[  1-\frac{A_2^{ 2(\delta -1)} }{C^2}
\, \dot{a}^2 \tilde{r}^2 \right]
-\frac{A_2^{\frac{-q+2\delta-1}{2}} }{C^2} \, 
\dot{a}  \tilde{r} \right\}
d\bar{t}dR \nonumber\\
&&\nonumber\\
& \, & + \left\{-A_2 
\left[ 1-\frac{A_2^{ 2(\delta -1)} }{C^2}\, 
\dot{a}^2  \tilde{r}^2 \right]\beta^2 
+\frac{2A_2^{\frac{-q+2\delta -1}{2}}}{C^2}\, 
\dot{a}\tilde{r}\beta +\frac{1}{A_2^qC^2} \right\} 
dR^2 \nonumber\\
&&\nonumber\\ 
& \, & + R^2d\Omega^2_{(2)} 
\,.\label{15}
\end{eqnarray}
By choosing the function $\beta$ as 
\begin{equation}\label{16}
\beta= \frac{ A_2^{ \frac{-q+2\delta-3}{2} } }{C^2}\, 
  \frac{\dot{a} \, \tilde{r}}{  1-
\frac{A_2^{2(\delta -1)} }{C^2}\, 
\dot{a}^2 \tilde{r}^2 }  
\end{equation}
the cross-term is eliminated and the metric becomes
\begin{eqnarray}
ds^2 & = & -A_2 D F^2 d\bar{t}^2 +\frac{1}{A_2^q C^2} \left[
1+\frac{ A_2^{-q-1} H^2 
R^2}{C^2 D} 
\right] dR^2 \nonumber\\
&&\nonumber\\
&\, & + R^2 d\Omega^2_{(2)} \,, \label{18}
\end{eqnarray}
where $ H\equiv \dot{a}/a$ and 
\begin{equation}\label{17}
D\equiv 1-\frac{ A_2^{2(\delta -1)} }{C^2}\, 
\dot{a}^2\tilde{r}^2 =
1-  \frac{A_2^{-q-1} }{C^2}\, 
H^2 R^2 \,.
\end{equation}
Using the second of these equations, the metric~(\ref{18}) becomes
\begin{equation}
ds^2 = -A_2 D F^2 d\bar{t}^2 +\frac{dR^2}{A_2^q C^2 D} 
+ R^2 d\Omega^2_{(2)} \,.
\end{equation}
The equation  $g^{RR}=0$ locating the apparent horizons
is $  A_2^q C^2 D =0$ 
and  $ A_2^q \left( C^2 -H^2R^2 A_2^{-q-1} \right)=0 $. 
Apparent horizons exist if $A_2=0$ or $H^2R^2=C^2 
A_2^{q+1}$.  $A_2$ vanishes at  $ r=  C_2 $, which 
describes the  Schwarzschild  horizon in the GR limit  
$\delta\rightarrow  0$. This is a spacetime singularity where
 the Ricci 
scalar $  {R^c}_c=\frac{ 6\left( \dot{H}+2H^2 \right)}{A_2(r)} $ 
diverges.

In the second case $H^2 R^2=C^2 
A_2^{q+1}$ we have 
\begin{equation} \label{24}
H R =\pm\left[ 1+\frac{2(q+2\delta -1)}{q}\, \frac{C_2 
a}{R}\, 
A_2^{\frac{2\delta-1-q}{2}} \right] A_2^{\frac{q+1}{2}} \,,
\end{equation}
choosing the positive sign for an expanding 
universe. When $\delta \rightarrow 0$, this equation reduces 
to $ HR =  
\left[ 
1+\frac{2\delta C_2 a}{R}\, A_2^{-\left(1-\frac{3\delta}{2} 
\right)} \right] A_2^{1-\delta} $. 

Two limits are now interesting:  the first one is  
$C_2\rightarrow  0$, in which the central object disappears 
leaving FLRW space, $r=\tilde{r}$  and $ R$  
reduce to the comoving and the areal radius, respectively, and  
eq.~(\ref{24}) reduces to $ 
R_{c}=1/H$, the radius of the cosmological horizon.  The second limit
of interest is $\delta\rightarrow 0 $:  the theory now degenerates into GR 
and eq.~(\ref{24}) reduces to $A_2=0$  or $r=C_2$ with $H\equiv 0$.

Eqs.~(\ref{3again}) and (\ref{8again}) allow one to express the left hand side of 
eq.~(\ref{24}) as
\begin{equation}
HR=\frac{ \delta \left( 1+2\delta \right)}{1-\delta}\, t^{ 
\frac{2\delta^2+2\delta -1}{1-\delta}} \, \frac{C_2}{x} \, \frac{ 
\left( 1-x \right)^{ \frac{ q+2\delta-1}{q}}}{\left( 1+x\right)^{ 
\frac{-q+2\delta-1}{q}}} \,,
\end{equation}
where $x \equiv C_2/r$. The right hand side of~(\ref{24})  is
\begin{equation}
\left( \frac{1-x}{1+x} \right)^{\frac{q+1}{q} }  \left[ 
1+\frac{2\left( 
q+2\delta-1\right)}{q} \, \frac{x}{\left( 1-x\right)^2} \right] 
\,,
\end{equation}
and eq.~(\ref{24}) is simply
\begin{eqnarray}
 \frac{1}{t^{ \frac{1-2\delta-2\delta^2}{1-\delta} } } 
& = & \frac{\left( 1-\delta\right)}{\delta\left( 
1+2\delta\right)C_2}\, 
\frac{ x \left( 1+x\right)^{ \frac{ -2q+2\delta-2}{q} }}{ \left( 
1-x\right)^{\frac{2\left( \delta-1\right)}{q}}} \nonumber\\
&&\nonumber\\
& \cdot & 
\left[ 1+\frac{2\left( q+2\delta -1 \right)}{q}\, 
\frac{x}{(1-x)^2} \right] \label{DELTA}
\end{eqnarray}
Here $\frac{1-2\delta- 2\delta^2}{1-\delta}$ is positive 
for $0<\delta < \frac{\sqrt{3} \, -1}{2} \simeq 0.366$. 
The radii $R$ of the apparent horizons and the time $t$  
can  be expressed in the parametric form 
\begin{eqnarray}
R (x) &=& t(x)^{ \frac{\delta \left(1+2\delta 
\right)}{1-\delta}}  \frac{C_2}{x} \, \left( 
1-x\right)^{\frac{q+2\delta-1}{q}} \left( 
1+x \right)^{\frac{q-2\delta+1}{q}} \,,\\
&&\nonumber\\
t(x) &=&   \left\{ \frac{\left( 1-\delta\right)}{\delta\left( 
1+2\delta\right)C_2}\, 
\frac{ x \left( 1+x \right)^{ \frac{ 2\left( 
-q+\delta-1\right)}{q} }}{ 
\left(  1-x\right)^{\frac{2\left( \delta-1\right)}{q}}}
\left[ 1+\frac{2\left( q+2\delta -1 \right)x}{q(1-x)^2} \right] 
\right\}^{\frac{1-\delta}{2\delta^2+2\delta-1}} \,,
\label{questa} 
\end{eqnarray}
using  $x$ as  a parameter. 
This parametric
representation of the horizons radii produces the same ``S-curve'' 
phenomenology of the Husain-Martinez-Nu\~nez solution \cite{myClifton}.

\subsection{Other solutions}

Few other solutions which, judging from their apparent horizons, can be 
interpreted as black holes embedded in cosmological backgrounds are 
known in Brans-Dicke \cite{SakaiBarrow, SakaiBarrow2} and in other 
theories of gravity (see, {\em e.g.}, those of Ref.~\cite{NozawaMaeda08} 
in Einstein-Gauss-Bonnet gravity, of Ref.~\cite{Charmousis} in higher 
order gravity, and of Ref.~\cite{MaedaWillisonRay11} in Lovelock gravity, 
and brane-world model solutions are known). Einstein-Gauss-Bonnet and 
Lovelock gravity, in particular, have been studied but they are 
appropriate in dimension $D>4$ and here we have restricted ourselves to 
$D=4$ --- the zoo of black objects (Myers-Perry black holes, black 
strings, black rings, black Saturns, {\em etc.}) is much larger in 
higher dimension \cite{HighDbook}. Moreover, most of the analytical 
solutions known are static or stationary. We did not mention here 
the variety of stringy and supergravity black holes (although 
most of them are stationary), which also deserve attention.

A problem is that, when trying to identify the physical mass of a 
non-asymptotically flat black hole in GR, we have made use of the 
Misner-Sharp-Hernandez mass and of the Kodama vector, defined in GR and 
for spherical symmetry. Thus far, beyond GR, the Misner-Sharp-Hernandez 
mass has been extended only to Einstein-Gauss-Bonnet gravity 
\cite{MaedaNozawa} (see \cite{Caietal09} for an attempt to define it in 
cosmological spaces in $f\left({R^c}_c\right)$ gravity) and is not 
available in alternative theories of gravity.

\section{Conclusions}

In this short review we have considered black holes with evolving horizons 
and, necessarily, we have given up the concept of event horizon and we have 
adopted the apparent and trapping horizon as its replacement. At present, 
this concept seems the best replacement and is widely used in practical 
(numerical) investigations, but it suffers from the drawbacks of being 
defined in a foliation-dependent way and of possibly becoming a timelike 
surface. The validity of the thermodynamics of apparent/trapping horizons 
needs to be studied better, as it seems that a quantum generalized second 
law does not hold for apparent and trapping horizons, while it holds for 
causal horizons, at least for $1+1$ dilaton gravity (to which GR reduces in 
spherical symmetry) and for conformal vacua and coherent states 
\cite{Wall12}.

We have focused on spacetimes 
representing, at least for part of their temporal extent, black holes 
embedded in cosmological backgrounds, of which a few solutions are 
known. We have considered only spherically symmetric spacetimes in $D=4$ 
dimensions, with the goal to provide relatively simple explicit 
analytical examples that could be used as toy models for various 
investigations of the thermodynamics of apparent/trapping horizons, 
spherical accretion \cite{Babichevetal04, ChenJing05, IzquierdoPavon06, 
PachecoHorvath07, MaedaHaradaCarr08, GaoChenFaraoniShen08, 
Guarientoetal08, 
Guariento2, Guariento3, Guariento4, Guariento5, Guariento6, 
Sun08, Sun09, GonzalezGuzman09, HeWangWuLin09} (possibly of 
interest for primordial black holes), quantization of black hole areas 
\cite{VisserBHareas1, VisserBHareas2, BHareas}, the issue of 
cosmological expansion versus local dynamics \cite{CarreraGiuliniRMD10}, 
and other theoretical topics. In addition, these solutions are of 
interest in themselves and, after all, they are not so simple. Even in 
the context of GR, where the McVittie spacetime has been known since 
1933, this type of solution is poorly understood.

When asymptotic flatness is given up and non-stationary matter is 
allowed outside the black hole, there is no Jebsen-Birkhoff theorem and 
there is no ``general'' solution of the Einstein equations, as is 
instead the case for vacuum asymptotically flat solutions of GR (for 
this case, it is well known from a variety of studies that the 
Kerr-Newman black hole is the most general 
stationary, axisymmetric, asymptotically flat, vacuum solution 
\cite{FrolovNovikov, 
Waldbook, Poissonbook}). The analytical solutions that can be given 
explicitly are probably pathological in this sense, rather than being 
generic (in some mathematical sense to be specified). The rather bizarre 
phenomenology of the apparent horizons in the solutions examined begs 
the question of whether they are even physically meaningful. They host 
naked singularities for part of the time and it is known that naked 
singularities form during  gravitational collapse, but 
they do not seem to be generic and ``typical'' choices of initial data 
result in black holes rather than naked singularities \cite{Christodolou94, 
Christodolou99}. It could well be 
that the solutions examined are exceptional rather than typical in the 
landscape of possible solutions.  The ``comoving mass'' solution is a 
late-time attractor in the generalized McVittie class, but it is not a 
generic cosmological black hole solution with spherical symmetry 
\cite{CarreraGiuliniRMD10, CarreraGiuliniPRD10}.  
No definitive statement can be made at 
present. Moreover, spacetimes hosting naked singularities cannot be 
obtained as the development of regular Cauchy data. Nevertheless, these 
spacetimes with time-evolving horizons tell us something about the 
theory of GR which may be important when trying to move beyond the 
paradigm of stationary and asymptotically flat vacuum black holes which 
has characterized black hole research thus far.

Things become even more uncertain when moving beyond the context of GR. 
To compound our ignorance, sometimes solutions of the field equations 
are discovered but no attempt is made to interpret them. Until not long 
ago, one had to be almost apologetic about working outside of GR, but 
things have changed and nowadays relevance for alternative theories of 
gravity is often used as a justification, or seen as an added value, for 
theoretical work. It is recognized that Einstein gravity will fail at 
some point and research in alternative theories is timely and 
important. We have only a handful of exact cosmological black holes in 
alternative gravity and we can only take a glimpse into this unexplored area. 
Some of the phenomenology of apparent horizons found in GR repeats 
itself in scalar-tensor gravity, but preliminary research has unveiled a 
wider range of behaviours.

At the moment, one could think of classifying cosmological black hole 
solutions of the field equations of a theory of gravity in two ways:

\begin{enumerate}

\item based on the type of matter filling the background FLRW universe 
({\em e.g.} dust, general perfect fluid, imperfect fluid, or scalar field);

\item based on the technique used to generate the solution from a known ``seed'' 
metric (when applicable), for example conformally transforming the Schwarzschild or Kerr 
solution in some coordinate system \cite{SultanaDyer05}, 
or applying a Kerr-Schild transformation \cite{Vaidya77, PatelTrivedi82, McClurethesis};

\item based on the phenomenology of the apparent horizons.

\end{enumerate}

GR solutions with a perfect fluid include the (generalized) McVittie 
class (and its special case, the Schwarzschild-(anti-)de Sitter black 
hole). Solutions with a scalar field as a source include the 
Husain-Martinez-Nu\~nez, and the (generalized) Fonarev 
solutions. The phenomenology of apparent horizons distinguishes at least 
between the McVittie/Lema\^itre-Tolman-Bondi type with two 
appearing/disappearing (one black hole and one cosmological) apparent 
horizon, the Husain-Martinez-Nu\~nez/Clifton  
type with its ``S-curve''  
phenomenology \cite{HusainMartinezNunez, myClifton}, 
and Lema\^itre-Tolman-Bondi spaces with a ``double S-curve'' \cite{BoothBritsGonzalezVDB},  
but other behaviours 
appear in the Clifton-Mota-Barrow solutions of Brans-Dicke theory due to 
the wider range of parameters.

In general, in both GR and alternative gravities, it is rare to find an 
explicit analytical expression of the proper (areal) radius of an 
apparent horizon of a dynamical 
cosmological black hole solution which could be used, for example, to 
investigate its thermodynamics. To the best of our knowledge such an expression 
is available only for the ``comoving mass'' subclass of the generalized 
McVittie class of solutions of GR. For other analytical solutions of the field 
equations the apparent horizon can only be obtained numerically or is given by 
an implicit analytical expression which is not very useful in practice.

The black hole solutions considered here represent eternal black holes 
which have not been created in a collapse process, but have existed 
forever or are created together with the universe in the Big Bang.

We can identify several other open problems: one is the possible 
generalization of the Misner-Sharp-Hernandez mass to non-spherical 
systems in GR, and its extension to alternative theories of gravity. 
Since the Misner-Sharp-Hernandez mass seems to be intimately connected 
with the Kodama vector, the problem can be seen as generalizing this 
vector to non-spherical systems and beyond GR. Another important problem 
concerns the thermodynamics of apparent/trapping horizons. There is a 
fairly large literature on this subject (both for black hole and 
cosmological horizons), which we cannot review here, and the 
``tunneling'' method has been applied to the computation of the Hawking 
temperature of apparent horizons, including time-dependent ones (see the 
review \cite{VanzoAcquavivaDiCriscienzo11}). However, contrary to the 
case of stationary black holes in which several independent calculational  
methods are available and produce the same result, for time-dependent 
apparent horizons only the Parikh-Wilczek ``tunneling'' method seem 
to deliver  
results and it is necessary to confirm these results with independent 
methods of calculation.

To conclude, cosmological black hole spacetimes as examples of 
time-evolving black holes disclose some of the complications of gravity 
and exhibit puzzling phenomenology. They can be useful for various 
areas of research in gravitational physics and quantum field theory in 
curved space but, unfortunately, we know too little about them. It is 
auspicable that in the near future more research is devoted to 
obtaining new solutions of this kind and, above all, understanding their 
physics.

\section*{Acknowledgments} 
It is a pleasure to thank Alex Nielsen, Hideki Maeda, Ivan Booth, John 
Barrow, Enzo Vitagliano, Thomas Sotiriou, Kayll Lake, Changjun Gao, 
Daniel Guariento, Matt Visser, Stefano Liberati, Sergio Zerbini, and 
Andres Zambrano Moreno for conversations or comments over the years (and 
Andres also for drawing figs.~\ref{McV1} and~\ref{McV2}), and Luciano 
Vanzo for his invitation to contribute to this volume.  This work is 
supported by the Natural Sciences and Engineering Research Council of 
Canada and by Bishop's University.

\end{document}